\documentclass[iop]{emulateapj}

\usepackage{amsmath}
\usepackage{longtable}

\usepackage{natbib}
\usepackage{color}
\usepackage{hyperref}
\hypersetup{colorlinks=true,urlcolor=black,linkcolor=red,citecolor=blue}


\newcommand{\mpccm}{\ensuremath{m_\mathrm{p}\,\mathrm{cm}^{-3}}}

\shorttitle{AGN Outflow Winds}
\shortauthors{Dugan et al.}

\begin{document}

\title{AGN Outflow Shocks on Bonnor-Ebert Spheres}
\author{Zachary Dugan\altaffilmark{1}, Volker Gaibler\altaffilmark{2}, Rebekka Bieri\altaffilmark{5}, Joseph Silk\altaffilmark{1,3,4}, Mubdi Rahman\altaffilmark{1}}
\altaffiltext{1}{The Johns Hopkins University Department of Physics \& Astronomy, Bloomberg Center for Physics and Astronomy, Room 366
3400 N. Charles Street, Baltimore, MD 21218, USA}
\altaffiltext{2}{Universit\"at Heidelberg, Zentrum f\"ur Astronomie, Institut f\"ur Theoretische Astrophysik, Albert-Ueberle-Str. 2, 69120 Heidelberg, Germany}
\altaffiltext{3}{Institut d'Astrophysique de Paris, UMR 7095, CNRS, UPMC Univ. Paris VI, 98 bis Boulevard Arago, 75014 Paris, France}
\altaffiltext{4}{Beecroft Institute for Cosmology and Particle Astrophysics, University of Oxford, Keble Road, Oxford OX1 3RH, UK}


\begin{abstract} 
Feedback from Active Galactic Nuclei (AGN) and subsequent jet cocoons and outflow bubbles can have a significant impact on star formation in the host galaxy.  To investigate feedback physics on small scales, we perform hydrodynamic simulations of realistically fast AGN winds striking Bonnor-Ebert (BE) spheres and examine gravitational collapse and ablation.  We test AGN wind velocities ranging from 300--3,000 km s$^{-1}$ and wind densities ranging from 0.5--10 \mpccm.  We include heating and cooling of low- and high-temperature gas, self-gravity, and spatially correlated perturbations in the shock, with a maximum resolution of 0.01 pc.  We find that the ram pressure is the most important factor that determines the fate of the cloud.  High ram  pressure winds increase fragmentation and decrease the star formation rate, but also cause star formation to occur on a much shorter time scale and with increased velocities of the newly formed stars.  We find a threshold ram pressure of $\sim 2\times10^{-8}$ dyne cm$^{-2}$ above which stars are not formed because the resulting clumps have internal velocities large enough to prevent collapse.  Our results indicate that simultaneous positive and negative feedback will be possible in a single galaxy as AGN wind parameters will vary with location within a galaxy.  
\end{abstract}

\section{Introduction}
The role Active Galactic Nuclei (AGN) play in the evolution of their host galaxy remains an outstanding problem in theoretical astrophysics \citep{Silk98}.  The discrepancy between the number of luminous galaxies predicted in a $\Lambda$CDM universe and observed by astronomers persists and is mainly explained by negative feedback through AGN quenching \citep{Weinmann}.  In many of today's cosmological simulations, this feedback is treated through the injection of thermal energy into the galaxy \citep[e.g.][]{Springel,Sijacki}, others actually include jets and outflows \citep{Kimm12, Dubois10}, and both contribute to achieve the necessary quenching to match with the observed luminosity functions.  

Observers still investigate the different impacts of jets and outflows on the host galaxy.  Recent observational evidence shows greater star formation rates (SFR) in radio-loud quasars than in radio-quiet quasars.  \citet{Kalf2012} analyze approximately 20,000 quasars from the Sloan Digital Sky Survey (SDSS), and employ the [OII] emission line to calculate SFRs.  They find higher SFRs in the radio-loud AGN, indicating AGN jet triggered star formation.  \citet{Kalf2014} utilize \textit{Herschel}-ATLAS \citep{Eales2010} and find that the far-infrared (FIR) data are critical to calculating SFRs in AGN, particularly in quasars where radiation from the accretion disk can contaminate calculations of the SFRs calculated from optical and ultraviolet (UV) data.  Once the FIR data is incorporated into the SFR calculations, the group finds both that high redshift radio loud AGN can have high star formation rates and that these bursts of star formation happen simultaneously with the growth of Super Massive Black Holes (SMBH).  These results are in tension with the idea of universal AGN quenching.  Similarly, \citet{Zinn13} study several hundred AGN from the Chandra Deep Field South and compare the SFRs, as determined from archived infrared data, finding higher SFRs in the radio-loud AGN, concluding AGN jets trigger star formation.  The differences in SFRs between radio-loud and radio-quiet quasars underscore the importance of understanding the feedback mechanisms on a smaller scale.  

However, with all the restrictions inherent to cosmological simulations on large physical scales, many of the critical physical processes in AGN feedback cannot be resolved due to density, velocity, and related time scales and dynamic ranges.  \citet{Gaibler12} simulate AGN jet feedback on a smaller, single-galaxy scale and show that mildly relativistic jets may induce star formation in their host galaxies.  The discrepancy between the two approaches requires a closer look on smaller time and physical scales to treat the problem.  

\citet{Wagner15} also explore both positive and negative AGN feedback from a theoretical perspective and identify the challenges of simulations.  One such problem is the huge range of physical scales important to AGN feedback, scaling from the pc size of the accretion disc, out to tens of kpc to the edges of the host galaxies, extending even further to Mpc scales and interaction with the inter-galactic medium.  Covering these spatial scales with good resolution while simultaneously capturing the extreme velocities close to the speed of light is virtually impossible.  Additionally including self-gravity and detailed radiative processes would make this computationally prohibitive.  It is important then to focus on specific aspects of AGN feedback with a given suite of simulations.  As an example, \citet{Bieri15} examine the feedback effect from the pressure confinement of a jet driven bubble on a disk radio galaxy, avoiding the computational costs of an actual jet by artificially introducing overpressure.  This frees up computational power to include self gravity.  They find the pressure causes an increased fragmentation of dense clouds in the host galaxy and subsequent increase in star formation (positive feedback).  

Another specific aspect of AGN feedback that is not well understood is the impact of winds from expanding jet cocoons or outflow bubbles on dense clouds of gas on small scales (i.e. cloud scales), though there is some observational evidence of this.  \citet{Tremblay16} analyze observations from ALMA and find that AGN jets can blow giant molecular clouds away from the galactic nucleus with jet-driven bubbles before gravity draws them back, similar to mechanical pumps.  These observations reveal star formation along the outer edges of the clouds that is potentially triggered by the expansion of the jet bubble.  Employing observations from the Measuring Active Galactic Nuclei Under MUSE Microscope (MAGNUM) survey on NGC 5643, a radio-quiet AGN with outflows, \citet{Cresci15} find positive feedback.  The observations show star formation in clumps within the double sided ionization cones with high-velocity gas, likely caused by the compression of the clouds from the outflow.  They also find a ring of star formation at 2.3 kpc, possibly triggered by the AGN outflow, a phenomenon also seen in \citet{Gaibler12} and \citet{Dugan14}.  

Simulations of AGN feedback on smaller scales provide a complementary view to observations as well as to large scale simulations.  The fate of the clouds and hence potential star formation is eventually decided on small scales.
\citet{Mellema02}, \citet{Cooper09} and \citet{Zubovas_14_out} have performed important studies to this end. Understanding the relationship between these winds and the clouds of gas available to form stars in a galaxy is critical to understanding the role AGN feedback has in galaxy evolution.

In this paper, we seek to build on past research of AGN feedback on clouds of gas by focusing on realistic AGN winds and gravitational collapse on the scale of a single cloud.  We simulate high velocity winds characteristic of AGN jet cocoons and outflow bubbles striking Bonnor--Ebert spheres \citep{Bonnor,Ebert}.  We have better resolution than in the past, and we include gravity, heating and cooling for the entire temperature range (including very low temperatures).

We organize this paper as follows: in Section 2, we discuss some previous literature, in Section 3 we describe our simulation setup, and in Section 4 our results.  In Section 5, we include our findings and conclude in Section 6.  

\section{Past Research}

A large number of cloud-crushing simulations have been performed in the past (of which we can only mention a selection), but few have employed values for the pressure, velocity, and density of the incoming shock wave that are realistic for AGN feedback as well as some crucial physical processes.  Adiabatic cloud crushing simulations may be scaled to the velocities and densities relevant for AGN feedback.  However, this is not possible when non-linear processes like gravitational collapse or cooling are included and have a significant impact.  Many of the simulations in the literature have not included gravity, so exploring its stabilizing effect, subsequent collapse, and potential star formation has been impossible for them.  We seek to include both realistic AGN wind parameters as well as gravity in order to probe whether or not a wind resulting from an AGN jet cocoon or an AGN wind bubble can cause the collapse of the ISM's dense clouds of gas and form stars.  

Few studies in the past have specifically focused on the impact of radio jet cocoons on a cloud of gas. \citet{Mellema02} investigate this topic through 2D simulations.  They list three main phases of wind--cloud interaction, first described by \citet{Klein94}.  In the first phase, the blast wave overtakes and interacts with the cloud.  In the second, the cloud is compressed through ram pressure and possibly thermal pressure if the cloud is under-pressured with respect to the environment.  This phase lasts for the cloud crushing time:
\begin{equation}
t_\mathrm{cc} = \chi^{1/2}R_\mathrm{c}/v_\mathrm{w}
\end{equation}
where $\rho_\mathrm{c}$ and $\rho_\mathrm{w}$ are the cloud and wind densities, respectively, $\chi=\rho_\mathrm{c}/\rho_\mathrm{w}$, $R_\mathrm{c}$ is the cloud radius, and $v_\mathrm{w}$ the wind velocity.  In the third phase, shocks from different sides of the cloud collide, forming a rarefaction wave back through the shocked gas, a process that typically takes a few cloud crushing times.  

\citet{Mellema02} use initial values for the cloud of $\rho=10\,\mpccm$, temperature $10^4$ K, $M=9.5\times10^5 M_\odot$, and size of 200 pc, and for the wind they chose $v_\mathrm{w}=3,500$ km s$^{-1}$.  Their cloud parameters are meant to resemble a Giant Molecular Cloud (GMC).  They show that with radiative cooling, the cooling time for the cloud is substantially shorter than both the cloud crushing time and the Kelvin-Helmholtz time:
\begin{equation}
t_\mathrm{KH} = R_\mathrm{c}(\rho_\mathrm{c}+\rho_\mathrm{w})/(v_\mathrm{c}-v_\mathrm{w})(\rho_\mathrm{c}\rho_\mathrm{w})^{1/2}
\end{equation}
where $v_\mathrm{c}$ is the cloud velocity.  They simulate uniform gas distributions in the shape of a circle and an ellipse, and find that at the conclusion of both simulations, less than $1\%$ of the original cloud has mixed in with the ambient gas.  This indicates the slow evaporation of the cloud under the conditions of a radio jet cocoon.  They also find that the radiative cooling occurs quickly enough that most of the cloud collapses into dense filaments that are likely to form stars, agreeing with the scenario of jet induced star formation.  However, self-gravity is not simulated, and star formation is not explicitly calculated.  

Some simulations of shocks resulting from other astrophysical phenomena cover parameter spaces pertinent to AGN feedback.  For example, \citet{Orlando05} simulate a supernova shock wave passing over a small uniform spherical gas cloud and find that thermal conduction and radiative cooling are critical to cloud survival.  They use two different scenarios, each with different velocities and temperatures, wind velocities of 250 and 430 km s$^{-1}$, densities of 0.4 $\mpccm$ in both, and temperatures of $1.6\times10^6$ and $4.7\times10^6$ K.   The sphere had a density of 1 $\mpccm$ and a temperature of $10^3$ K, while the ISM had a density of 0.1 $\mpccm$ and a temperature of $10^4$ K.  They show the importance of thermal conduction in suppressing destructive instabilities in the first case, though the cloud does expand and eventually evaporate.  In the second case they show that the radiative cooling is pivotal to the formation of cold, dense clumps after the wind passes over the cloud.  Gravity is not included, and no star formation is calculated.  

Also relevant to AGN feedback, \citet{Cooper09} explore shocks from starbursts on fractal clouds and uniform spheres alike and also conclude on the importance of radiative cooling.  They use  wind values of $\rho = 0.1$ $\mpccm$, $v = 1.2 \times 10^5$ cm s$^{-1}$, $T = 5 \times 10^6$ K, and sphere values of $r = 5$ pc, $M = 523 M_\odot$, $\bar{\rho}=91$ $\mpccm$, $T = 5\times 10^3$ K.  The maximum resolution is 0.13 pc.  
The winds form dense filamentary structures like those observed in starburst driven winds.  They also find that radiative cooling is essential to the longer survival time of the clouds and filaments.  The wind does not heat or accelerate radiative clouds as much as their adiabatic counterparts, decreasing the impact of Kelvin-Helmholtz instabilities and increasing the lifetime of the clouds and filaments.  The geometry of the cloud does impact their results and the wind destroys less dense regions more quickly.  While they indicate that self-gravity may cause the filaments to collapse, they do not include gravity or look at subsequent star formation in this study.  

\citet{Pittard10} perform 2D cloud crushing simulations with velocities relevant to AGN feedback and find that KH instabilities dominate in the cloud's destruction.  They assume values of $r = 2$ pc, central density $\rho_\mathrm{c} = 60$ $\mpccm$, and temperature of 8000 K, and is in pressure equilibrium with the surrounding gas of density $\rho_a = 0.24$ $\mpccm$  and temperature T$_a = 8 \times 10^6$ K.  Cloud density follows the profile $\rho(r)=\rho_{amb}[\Psi+(1-\Psi)\times\frac{1}{2}\left(1+\frac{\alpha-1}{\alpha+1}\right)$ where $\rho_{amb}$ is the ambient density, $\Psi$ is a parameter selected to create a specific density contrast between the center and edge of the sphere, $\alpha=\exp[\mathrm{min}(20, 10( (r/r_c)^2-1)]$, and $r_c$ is the cloud radius.  They simulate velocities of 650, 1,300, and 4,300 km s$^{-1}$.  They also find the formation of dense filaments in many of the simulations, and note that the presence of turbulence in the post  wind speeds up the destruction time of the cloud.  However, they too do not include gravity.

\citet{Zubovas_14_out} perform Smoothed Particle Hydrodynamics (SPH) simulations aimed at determining the role of AGN outflow or jet generated external pressure on a turbulent, spherically symmetric molecular clouds as well as fractal clouds.  They begin with the initial conditions of a massive cloud of $10^5 M_\odot$ with a 10 pc radius for an average particle density of 380 cm$^{-3}$.  The group simulates turbulent velocities 3--10 km s$^{-1}$ and winds of up to 300 km s$^{-1}$, and include gravity.  In the case of the shocked sphere, they employ a uniform density distribution for the cloud.  The ISM, which comprises the wind, has a density of $\rho\sim1$ $\mpccm$ and a temperature of either $T = 10^5$ or 10$^7$ K.  They conclude that the pressure confinement of the shock wave increases the star formation rate and efficiency, i.e. that a higher fraction of the molecular gas is converted into stars.  They also find that the pressure confinement decreases the sizes of the resulting star clusters, and that cloud rotation and shear against the ISM is not important unless the shear velocity is substantially greater than the sound speed of the constraining ISM.  Their results are consistent with positive AGN feedback in gas rich clouds with increased pressure \citep{Gaibler12, Bieri15}.

We summarize the paramters of these studies in Table \ref{tab:past_research}.

\begin{deluxetable*}{cccccccccc} 
\tablecaption{Past Research}
\tablewidth{0pc} 
\tablecolumns{10}
\tablehead{ \colhead{Study} & \colhead{$\rho_s$} & \colhead{$v_s$} & \colhead{P$_\mathrm{ram}$} & \colhead{M$_\mathrm{c}$} & \colhead{$\rho_c$} & \colhead{code} & \colhead{dim} & \colhead{Resolution} & \colhead{grav} \\
\colhead{} & \colhead{[$\mpccm$]} & \colhead{[km s$^{-1}$]} & \colhead{[dyne cm$^{-2}$]} & \colhead{[M$_\odot$]} & \colhead{[$\mpccm$]} & \colhead{} & \colhead{ } & \colhead{[pc]} & \colhead{} }
\startdata
\citet{Mellema02} & 1e-2 & 3,500 & 2.05e-09 & 9.5e5 & 10 & AMR & 2D & 0.5 & no \\  
\citet{Orlando05} & 0.4 & 250 - 430 & 4.2e-10 - 1.2e-9 & 0.1 & 1 & AMR & 3D & 1e-2 & no \\  
\citet{Cooper09} & 0.1 & 1,200 & 2.4e-9 & 700 - 1400 & 60-130 & AMR & 3D & 0.1-0.8 & no \\  
\citet{Pittard10}$^{(a)}$ & 0.24 & 650 - 4,300 & 1.7e-9 - 7.4e-8 & 50 & 60 & AMR & 2D & 0.02 & no\\ 
\citet{Zubovas_14_out} & 1 & 300 & 1.5e-9 & 1e5 & 380 & SPH & 3D & -- & yes \\ 
this study& 1-10 & 300-3,000 & 7.5e-10 - 4.5e-7& 72 & 300-900 & AMR & 3D & 1e-2 & yes 
\enddata
\tablenotetext{}{\begin{center}$^{(a)}$ Adiabatic and scalable.\\
\end{center}
}
\label{tab:past_research}
\end{deluxetable*}

\section{Simulations}

We seek to improve past research on wind cloud interactions by focusing on the high-velocity winds and shocks characteristic of AGN jet cocoons and outflow bubbles propagating through the host galaxy's ISM on a cloud sized scale.  We use Bonnor--Ebert (BE) spheres \citep{Bonnor,Ebert} in place of uniform-density spherical clouds, and our spheres have higher peak densities, more representative of dense clumps within the ISM.  We employ realistically high velocities in our winds to emulate those found from detailed models and simulations of AGN jets and winds.  We include self-gravity, spatially correlated velocity perturbations, and a model of cooling and heating processes for all gas phases involved, in contrast to previously employed simplifications.  

\subsection{Numerics and setup}

We use RAMSES \citep{Teyssier} for our simulations, a non-relativistic second-order Godunov-type shock-capturing hydrodynamics code with adaptive mesh refinement.  The computational domain is a cubic box of 10 pc on a side with a minimum and maximum effective resolution of $2^6$ and $2^{10}$ cells per side. We chose the HLLC Riemann solver, minmod slope limiter, an adiabatic index of $1.6666$, and active ``pressure-fix'' (enabling a hybrid scheme that avoids negative pressure in high Mach number regions) for the numerical integration.

Hydrodynamical boundaries are set to zero-gradient outflow except the wind inflow boundary on the ``left'' (referring to our projections in Figs.~\ref{fig:density_projections} and \ref{fig:density_projections_fast}), which is set to a perturbed inflow at an angle of $10^\circ$ to the $x$ coordiante axis.  The non-zero angle is chosen to prevent grid alignment artifacts.  Details on the perturbations of the inflow are given in Sect.~\ref{sec:perturbations}.  The Poisson equation is solved by a multigrid method with an iteration stopping criterion of $10^{-5}$ and periodic boundary conditions because gravitational effects are limited to the small high-density regions.  All gas inside the computational domain initially is at rest, but then overrun by wind.

Cells are adaptively refined to $1/300$ of the local Jeans length until reaching the maximum resolution (hence refining on the dense gas) and to 0.04 pc in all the wind (inflow) region upstream that could interact with the sphere in order to resolve the inflow perturbations, with straight injection for stability reasons. For most of the simulations, we track the cloud for 250 kyr (free-fall time at $4\times 10^4 \, \mpccm$).

\subsection{Bonnor--Ebert Spheres}

BE spheres are hydrostatic solutions of self-gravitating isothermal clouds embedded in a low-density environment.  In particular, their non-uniform density profiles and pressure balance with the ambient medium make BE spheres a more realistic description for dense clumps of molecular gas than uniform spheres, which are not stable gravitationally.  The Bok Globule provides an example of a hydrostatic blob of gas in pressure equilibrium with its environment and a density profile similar to that of a BE sphere \citep{Lada01}.  Because of their density profile and pressure balance, BE spheres are commonly employed as models to investigate low-mass star formation on small scales.  For low ratios ($<$ 14.1) of the sphere's central density to its outer-edge density, BE spheres are stable against small perturbations.  To assign values that would match those of a dense clump, we wanted to have a central and outer density a few orders of magnitude greater than the ambient medium.  To make collapse more difficult, we selected a ratio of central density to outer density to maximize the stability of the sphere, a value of 3.  We have the ambient medium initially at rest and with a density of $1\,\mpccm$, comparable to the wind densities of many of the simulations. The BE sphere has an outer-edge density of $300\,\mpccm$, and a central density of 900 $\mpccm$, greater than those in other simulations (see Table \ref{tab:past_research}).  We deem these high densities necessary to simulate the high density clumps within the ISM.  We select a temperature of 19 K, which results in a sound speed of approximately 0.4 m s$^{-1}$.  This choice of parameters results in an initial free-fall time of approximately 3 Myr and a cloud mass of 72 M$_\odot$.  

We derive our density profile below, beginning with the equation of pressure structure, equation of state, and equation of gravitational stability, and spherically symmetric Poisson equation of gravity:  
\begin{equation}
\frac{dP}{dr} = -G M(r)  \rho(r) / r^2
\end{equation}
 
\begin{equation}
P = \frac{\rho k_\mathrm{B} T}{\mu m_p} = \rho c_s^2
\end{equation}
 
\begin{equation}
\frac{dP}{dr} = -\frac{d\Phi}{dr} \rho(r)
\end{equation}
 
\begin{equation}
\frac{1}{r^2} \frac{\partial}{\partial r}( r^2 \Phi) = 4\pi G\rho
\end{equation} 

We can combine the equation of hydrostatic equilibrium and the ideal gas law before substituting it into the above equation:
\begin{equation}
-c_s^2 \frac{d}{dr} ln(\rho) = \frac{d\Phi}{dr}
\end{equation} 
 
 After substitution back into the spherically symmetric Poisson equation we get a form of the Lane-Embden Equation, a second order ordinary differential equation (ODE) that essentially defines a BE sphere:
\begin{equation}
\frac{1}{r^2} \frac{\partial}{\partial r}\left( r^2 \frac{\partial ln(\rho)}{\partial r}  \right) = 4\pi G\rho /c_s^2
\end{equation} 

To compute the density profile, one must define a central density, $\rho_0$, and then integrate numerically to the outer edge.  We also apply the boundary condition that $\Phi(r=0)=0$.  BE spheres are well-discussed in the literature, for more information see  \citet{Kaminski2014} and references within.

\subsection{Parameter Space}

We explore a parameter space that encompasses winds and shocks from starbursts to AGN jet cocoons and quasar outflows/winds.  Most important, our wind velocities range from 300--3,000 km s$^{-1}$ to properly simulate the impact AGN feedback could have on a BE sphere.  Most studies in the literature do not explore winds that exceed velocities of 1,200 km s$^{-1}$ despite galaxy scale simulations and observations that show wind velocities greater than 1,000 km s$^{-1}$, even when AGN are in quasar mode as opposed to radio mode.  While AGN jets and their cocoons may typically occupy the lower density \& higher velocity regime, quasar winds are expected to be more mass-loaded.

For a spherically symmetric galaxy, the energy conserving analytical models of a jet or outflow driven bubbles depend only on the power of the jet or outflow \citep{Wagner12_jet}.  The equations for the bubble's radius and resulting wind velocity are:
\begin{equation} \label{eq:bubble_radius}
R_\mathrm{b} = At^{3/5}
\end{equation}
\begin{equation} \label{eq:bubble_velocity}
v_\mathrm{w} = (3/5)At^{-2/5}
\end{equation}
where:
\begin{equation} 
A = \left(\frac{125P_\mathrm{jet}}{384\pi\rho_\mathrm{a}}\right)^{1/5}
\end{equation}
in which $P_\mathrm{jet}$ is the power of the jet or outflow, and $\rho_\mathrm{a}$ is the density of the ambient medium.  Using Equation \ref{eq:bubble_velocity}, we can calculate the bubble velocity and radius on full galaxy scales to demonstrate the range and locations of wind velocities from AGN feedback.  We also vary the density of the ambient medium from 0.5-10 $\mpccm$ to further demonstrate the range of resulting values.  We see that with that ambient density range, in the first few Myr, the bubble will have a radius less than 5 kpc and a velocity in the 1,000's of km s$^{-1}$.  After 20 Myr, the bubble will have a radius between 7 and 14 kpc and a velocity in the 100's of km s$^{-1}$.  Therefore, to understand the variable impact of AGN feedback, one must explore the full parameter space of AGN winds and the unique combination of high velocities in the 1,000's km s$^{-1}$ and densities resulting from black hole feedback.  

Also unlike previous simulations in the literature, our winds are highly pressurized, meant to mimic the highly pressurized environments inside AGN jet cocoons and outflow bubbles.  To be consistent with both theoretical and observational values for AGN winds, we simulate values within the ranges listed in \citet{Liu}, \citet{Gaibler12}, \citet{Wagner12_jet}, and \citet{Wagner13_out}.  We list the parameters for each of our simulations in Table \ref{tab:parameter_space}.  The run labels contain the density and velocity of the shock, so 'd1-v3000' would indicate wind density and velocity of 1 $\mpccm$ and 3000 km s$^{-1}$.  We simulate winds with the combinations of densities of 0.5, 1, 3, and 10 $\mpccm$ with velocities of 300, 1000, and 3,000 km s$^{-1}$.  The combination of a velocity of 3,000 km s$^{-1}$ and a density of 10 $\mpccm$ (i.e. even higher ram pressure) was not modeled. The lower density value of $0.5\, \mpccm$ was included particularly because thermal instability of the dense gas starts at $~ 1\,\mpccm$. Run 'd1-v1000-uni' used a uniform sphere with the same mass and radius as the BE sphere.
Additionally, the table includes the cloud crushing time and the Kelvin-Helmholtz timescale.

\begin{deluxetable*}{ccccccccccc} 
\tablecaption{Employed values for the simulations}
\tablewidth{0pc} 
\tablecolumns{11}
\tablehead{ 
\colhead{Run} & \colhead{$\rho$} & \colhead{v$_w^{(a)}$} & \colhead{v$_w$/c$^{(b)}$} & \colhead{P$_\mathrm{ram}$$^{(c)}$ } & \colhead{P$^{(d)}$} & \colhead{T$^{(e)}$ } & \colhead{$\chi$$^{(f)}$ } & \colhead{$t_\mathrm{cc}$$^{(g)}$} & \colhead{$t_\mathrm{KH}$$^{(h)}$} & \colhead{$\mathcal{M}$$^{(i)}$} \\
\colhead{Label} & \colhead{[$\mpccm$]} & \colhead{[km s$^{-1}$]} & \colhead{} & \colhead{[dyne cm$^{-2}$]} & \colhead{[dyne cm$^{-2}$] } & \colhead{[K]} & \colhead{ } & \colhead{[yr]} & \colhead{[yr]} & \colhead{} 
 } 
\startdata
d0-v300 & 0.5 & 300 & 0.001 & 7.52e-10 & 1e-09 & 8.70e+06  & 6e+02 & 95.84 & 166.09 & 1.07e+03 \\  
d1-v300 & 1.0 & 300 & 0.001 & 1.50e-09 & 1e-09 & 4.35e+06  & 3e+02 & 67.77 & 117.51 & 1.07e+03 \\  
d3-v300 & 3.0 & 300 & 0.001 & 4.51e-09 & 1e-09 & 1.45e+06 & 1e+02 & 39.13 & 68.00 & 1.07e+03 \\  
d0-v1000 & 0.5 & 1,000 & 0.0033 & 8.35e-09 & 1e-09 & 8.70e+06  & 6e+02 & 28.75 & 49.83 & 3.57e+03 \\  
d10-v300 & 10.0 & 300 & 0.001 & 1.50e-08 & 1e-09 & 4.35e+05  & 3e+01 & 64.29 & 112.59 & 1.07e+03 \\  
d1-v1000 & 1.0 & 1,000 & 0.0033 & 1.67e-08 & 1e-09 & 4.35e+06  & 3e+02 & 20.33 & 35.25 & 3.57e+03 \\
\hline 
d1-v1000-uni & 1.0 & 1,000 & 0.0033 & 1.67e-08 & 1e-09 & 4.35e+06  & 3e+02 & 20.33 & 24.54 & 3.57e+03 \\   
\hline
d3-v1000 & 3.0 & 1,000 & 0.0033 & 5.01e-08 & 1e-09 & 1.45e+06  & 1e+02 & 11.74 & 20.40 & 3.57e+03 \\  
d0-v3000 & 0.5 & 3,000 & 0.01 & 7.52e-08 & 1e-09 & 8.70e+06  & 6e+02 & 9.58 & 16.61 & 1.07e+04 \\  
d1-v3000 & 1.0 & 3,000 & 0.01 & 1.50e-07 & 1e-09 & 4.35e+06  & 3e+02 & 6.78 & 11.75 & 1.07e+04 \\  
d10-v1000 & 10.0 & 1,000 & 0.0033 & 1.67e-07 & 1e-09 & 4.35e+05  & 3e+01 & 6.43 & 11.26 & 3.57e+03 \\  
d3-v3000 & 3.0 & 3,000 & 0.01 & 4.51e-07 & 1e-09 & 1.45e+06  & 1e+02 & 3.91 & 6.80 & 1.07e+04   
\enddata 
\tablenotetext{}{\begin{center}Run d1-v1000-uni: uniform sphere with the same mass and radius as the BE sphere.\\
$^{(a)}$ Wind velocity.\\
$^{(b)}$ Wind velocity / c.\\
$^{(c)}$ Ram pressure of wind.\\
$^{(d)}$ Wind pressure.\\
$^{(e)}$ Wind temperature.\\
$^{(f)}$ $\chi=\rho_\mathrm{c}/\rho_\mathrm{w}$\\
$^{(g)}$ Cloud crushing time.\\
$^{(h)}$ Kelvin-Helmholtz time.\\ 
$^{(i)}$ Mach number- wind velocity relative to sound speed of sphere.  \\
\end{center}
}
\label{tab:parameter_space}
\end{deluxetable*}

We can also derive the acceleration and therefore the velocity of a spherical cloud by a bubble-driven wind:
\begin{equation} \label{eq:wind}
m_\mathrm{c} \frac{dv_\mathrm{c}}{dt}=C_D\rho_\mathrm{w}v_\mathrm{w}^2\times\pi R_\mathrm{c}^2
\end{equation}
in which $C_D$ is the drag coefficient, $\rho_\mathrm{b}$ is the density of the bubble, and $R_\mathrm{c}$ is the radius of the cloud.  We assume $C_D \sim 1$.  
Employing Equation \ref{eq:wind}, we can calculate the theoretical cloud velocity as a function of time using the values for the winds we list in Table \ref{tab:parameter_space} and the average values for the BE sphere.  This calculation results in expected velocities from 10-1000 km s$^{-1}$ on time scales less than 100 kyr. Yet this strongly depends on the (much changing) cloud parameters.

\subsection{Heating and Cooling}

Although the multi-phase nature of the interstellar medium with a huge range of densities and temperatures is well-established, hydrodynamic simulations on galaxy scales often only describe the warm and hot phases properly. The cooler parts of the ISM are difficult to model and widely mimicked by imposing a minimum pressure according to a polytropic equation of state or a minimum temperature.
And while resolution often makes this necessary, in particular with self-gravity, where the Jeans length needs to be well resolved to avoid numerical artifacts \citep{Truelove+1997}, this also means that the actual star-forming phase is never addressed and modelled explicitly and all star formation related results bear considerable uncertainty.

\begin{figure}
    \begin{center}
    \includegraphics[width=0.95\columnwidth,angle=0]{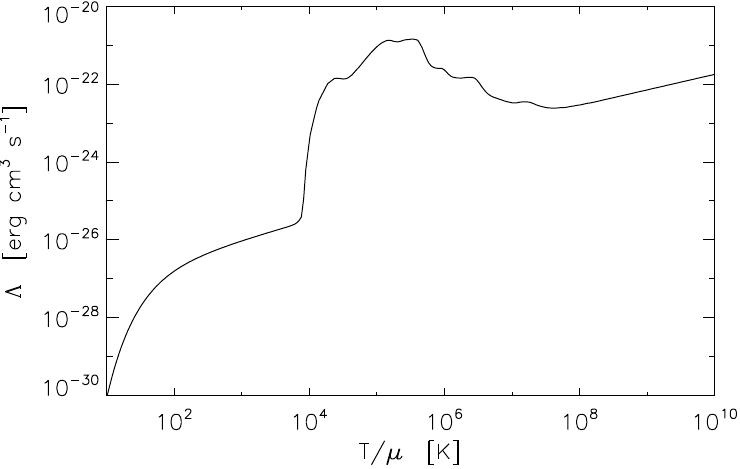}
    \end{center}
    \caption{The employed cooling function $\Lambda(T/\mu)$ combines the data of \citet{SutherlandDopita1993} with the low-temperature model of \citet{KoyamaInutsuka2002}.
    }
    \label{fig:cooling-lambda}
\end{figure}

\begin{figure}
    \begin{center}
    \includegraphics[width=0.95\columnwidth,angle=0]{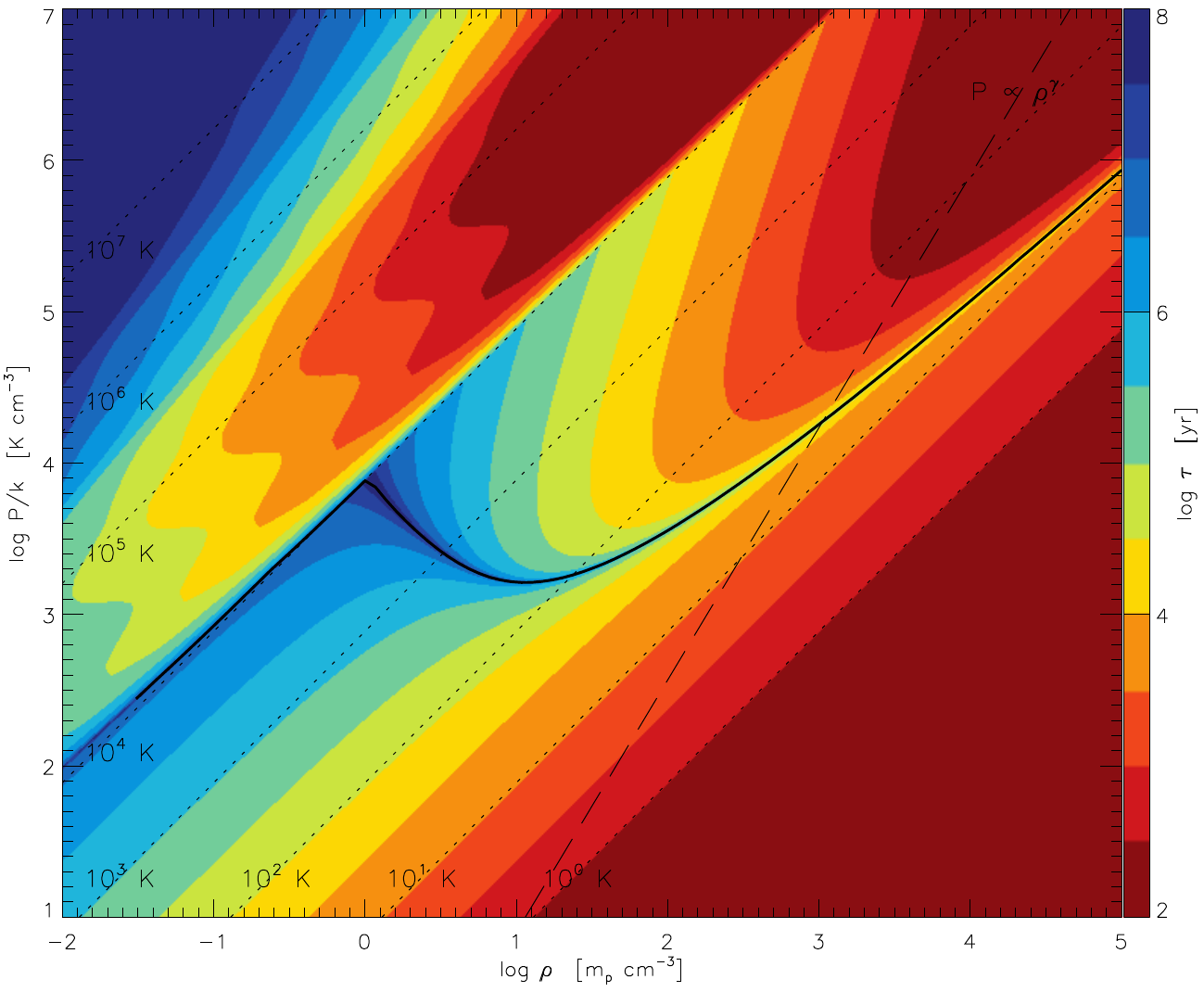}
    \end{center}
    \caption{Net heating/cooling time for the gas in $\rho$--$p$ space, as defined by $\tau = e / \dot{e}$ with the internal energy density $e$. The equilibrium curve is shown in solid black, constant temperature as dotted lines and the slope of an adiabatic process with $\gamma = 5/3$ as long-dashed line.
    }
    \label{fig:cooling-timescales}
\end{figure}

Since our study focuses on the pc scale, we necessarily need a thermodynamical model of the gas spanning the full range of temperatures, from $\sim 10$ K in cold ``molecular'' clouds to $>10^7$ K in the AGN wind. We handle this by adding a source term for heating and cooling to the energy equation with
\begin{equation}
    \frac{de}{dt} = \left( \frac{\rho}{m_\mathrm{p}} \right) \Gamma \, - \left( \frac{\rho}{m_\mathrm{p}} \right)^2 \Lambda(T/\mu) \: ,
\end{equation}
a heating parameter $\Gamma = 2 \times 10^{-26}$ erg s$^{-1}$ and a cooling function $\Lambda(T/\mu)$ for the entire temperature regime (Fig. \ref{fig:cooling-lambda}, where $\mu$ is the mean molecular weight in $m_\mathrm{p}$ at the respective temperature, $\mu \approx 2.3$ for molecular gas and $\mu \approx 0.62$ for the fully ionized gas). Numerically, the heating and cooling are computed with a linearized implicit method and substepping within the hydrodynamic time step. The high temperature domain ($T > 10^4$ K) is modeled after the collisional ionization equilibrium cooling function of \citet{SutherlandDopita1993} with a solar metallicity, the low temperature domain ($T < 10^4$ K) after \citet{KoyamaInutsuka2002} with their corrected fitting formula as given in \citet{VazquezSemadeni+2007}. For the transition between the two regimes, the high-temperature term of \citet{KoyamaInutsuka2002} is damped to continuously connect to and not exceed the \citet{SutherlandDopita1993} function. This has the additional benefit of avoiding the overestimated Lyman $\alpha$ cooling of the former and should bring us closer to the behavior found in detailed calculations of non-equilibrium chemistry and thermal balance \citep{Micic+2013}, while still being able to stick to simple one-fluid hydrodynamics. The \citet{KoyamaInutsuka2002} fit includes many low-temperature processes, has often been employed for converging flow studies without full chemistry and yields a two-phase behavior for the cold and warm neutral gas on the thermal equilibrium curve and thermal instability can drive equilibrium gas of $1 \, \mpccm$ to 240 $\mpccm$ (see Fig.~\ref{fig:cooling-timescales}). 

We define the cooling length as
\begin{equation}
L_{cool} = v_w \, \tau_\mathrm{cool}
\end{equation}
where $v_w$ is the wind velocity and $\tau_\mathrm{cool} = e / \dot{e}$ is the cooling time. 
For comparison, we calculate the cloud ablation time:
\begin{equation}
t_{abl} = R_C/v_w
\end{equation}
where $R_C$ is the cloud radius.  We include the ablation time, the cooling time, and the cooling length for each simulation in Table \ref{tab:time_parameter_space}.

While ideally the cooling length should be resolved, this is simply impossible for our setup and the maximum resolution of 0.01 pc, and hence we need to keep the implications of this in mind. When cooling shells are unresolved, cooling at the neigboring cells is overestimated, but for the shell itself is underestimated. As a result, the cloud could be more susceptible to Kelvin-Helmholtz and Raleigh-Jeans instabilities that ablate the clump.  The tendency for gravitational collapse might be weaker than if cooling shells were properly resolved.   However, with additional resolution, other physical factors such as thermal conduction or magnetic fields become important and drastically complicate required physics \citep[see also][]{GloverMacLow2007}. Predicting the impact of this lack of resolution is difficult and would require a much more sophisticated numerical model. However, these effects would mostly occur near the ``surface'' of the cloud and less affect the cloud cores and gravitational collapse.

\begin{deluxetable}{cccc} 
\tablewidth{0pc} 
\tablecolumns{4}
\tablecaption{Additional Time Scales for Simulation Parameters}
\tablehead{ \colhead{Run} & \colhead{${t_{abl}}^{(a)}$} & \colhead{${\tau_{cool}}^{(b)}$} & \colhead{${L_{cool}}^{(c)}$ } \\  \colhead{Label} & \colhead{[kyr]} & \colhead{[kyr]} & \colhead{[pc]} }
\startdata 
d0-v300  &  3.91e+00  &  1.14e+03  &  1.11e-05  \\ 
d1-v300  &  3.91e+00  &  5.71e+02  &  5.55e-06  \\ 
d1-v300  &  3.91e+00  &  1.06e+00  &  1.03e-08  \\ 
d0-v1000  &  1.17e+00  &  1.14e+03  &  3.70e-05  \\ 
d10-v300  &  1.17e+01  &  6.35e-03  &  2.06e-11  \\ 
d1-v1000  &  1.17e+00  &  5.71e+02  &  1.85e-05  \\
\hline 
d1-v1000-uni  &  1.17e+00  &  5.71e+02  &  1.85e-05  \\ 
\hline
d3-v1000  &  1.17e+00  &  1.06e+00  &  3.43e-08  \\ 
d0-v3000  &  3.91e-01  &  1.14e+03  &  1.11e-04  \\ 
d1-v3000  &  3.91e-01  &  5.71e+02  &  5.55e-05  \\ 
d10-v1000  &  1.17e+00  &  6.35e-03  &  2.06e-10  \\ 
d3-v3000  &  3.91e-01  &  1.06e+00  &  1.03e-07  
\enddata
\tablenotetext{}{\begin{center}Run d1-v1000-uni: uniform sphere with the same mass and radius as the BE sphere.\\
$^{(a)}$ Ablation time.\\
$^{(b)}$ Cooling time.\\ 
$^{(c)}$ Cooling length.\\
\end{center}
}
\label{tab:time_parameter_space}
\end{deluxetable}

\subsection{Spatially Correlated Perturbations}
\label{sec:perturbations}

While AGN blast waves and winds globally exhibit an approximately spherically symmetric expansion, they are not well-ordered flows, and show considerable perturbations on the smaller scales.  Overpressured, expanding jet cocoons of AGN jets show velocity components of propagation of the cocoons and their leading bow shocks through a multi-phase ISM \citep[Figure 9 of][]{Wagner12_jet}, and also velocity components of strong turbulence \citep[Figure 15 of][]{Gaibler+2009}.  The two together generate stochastic flow perturbations on the scales that are relevant for interaction of such winds with cold clouds. To mimic this effect, we use the spatially correlated perturbations generated by a superposition of comoving plane waves on top of a constant inflow velocity $\vec{v}_\mathrm{w}$,
\begin{equation}
    \vec{v} \: = \: \vec{v}_\mathrm{w}  +  \sum_\lambda \sum_i^{N_i} a_i \vec{n}_i \sin \left( k_\lambda \vec{n}_i \cdot \vec{x}^\prime + \phi_i \right) \, .
\end{equation}
where $k_\lambda$ are the $3$ discrete wave numbers considered (corresponding to wave lengths of $1/3$, 1 and 3 pc), $a_i$, $\vec{n}_i$ and $\phi_i$ are the amplitude, random direction and random phase of the $N_i$ waves at each wave number, respectively, and $\vec{x}^\prime = \vec{x} - \vec{v}_\mathrm{w} t$ is the location vector in the frame comoving with the AGN wind. We chose $N_i = 10$ plane waves at each wave number, plane wave amplitudes scaling $\propto k_\lambda^{-1/3}$ and the overall RMS amplitude of the velocity perturbations normalized to $20\%$ of the wind velocity $\vec{v}_\mathrm{w}$. Perturbations were only applied to the velocity variable on the boundary, but clearly propagate into the other flow variables inside the computational domain.  We note that these perturbations furthermore avoid artifacts from an artificially symmetric setup (e.g. during the compression of the cloud). With respect to the interaction of the wind with the cloud, the choice of wave numbers excites perturbations on scales smaller, similar to, and greater than the truncation radius of the BE sphere.

\subsection{Clump Analysis}

To better investigate star formation, we employ a clump finder to look at the individual dense clumps that form after the BE sphere is shocked, rather than look at the sphere as a whole.  This approach allows us to identify smaller pockets of gravitation instability and collapse that we would not see if we looked at the entire sphere as a single body, giving us a better idea of the trend of star formation.
We use the Ramses Clump Finder \citep{Teyssier14} to find and track dense clumps resulting from the wind and collapse of the BE sphere in the various simulations.  Typically clump finders employ either a cell based method, in which clumps and/or sink particles are formed based on conditions within an individual cell, or peak based conditions, in which clumps and/or sink particles are based on the hydrodynamic state surrounding density peaks.  This new Ramses Clump Finder uses a more sophisticated implementation of the peak based method, but with restrictions to exclude small density fluctuations, thus providing more realistic separations into individual clumps.  

This algorithm works in five main steps.  First, all cells with a density above a predetermined threshold are identified.  Next, local density maxima are are identified and given a peak identification number, and all the other cells are associated with one of these peaks based on steepest ascent.  Next, the saddle points between the previously identified peaks are identified, and are determined as significant or not based on the predetermined peak-to-saddle ratio.  All insignificant peak areas are merged with the closest significant ones, and the process is repeated until all the identified peaks are significant and finally reported as clumps for further investigation.  For more detail, see \citet{Teyssier14}.  We modify the standard clump finder so that instead of analyzing data on individual clumps we can analyze data on each cell within a clump.  Cell by cell analysis is particularly important for calculating the binding energy of each clump and subsequent collapse.  

To probe gravitational instability in the resulting clumps, we experimented with two physical parameters taken as inputs for the clump finder: the density threshold and the peak-to-saddle ratio.  The two values that best allow for the detection of gravitational instability are a threshold of 10$^{5.5}$ $\mpccm$ and a peak-to-saddle ratio of 1.2.  These values allow for a focus on the ultra-dense center of resulting clumps, precisely where collapse occurs and stars are formed.  We define our gravitationally unstable regions as those that are Jeans unstable and cannot support themselves with their own internal pressure, and that have binding energy in excess of their internal kinetic energies:
\begin{equation}
M_\mathrm{J} = 5 \langle R \rangle_m \langle P \rangle_m/(G\langle \rho \rangle_m) < M_\mathrm{clump}
\end{equation}
\begin{equation}
\sum_i \phi_i m_i > \sum_i m_i v_{{int}_i}^2 /2
\end{equation}
where $\phi$ is the gravitational potential, and we use the mass-weighted averages of radius, pressure, and density to calculate the Jeans mass.  We calculate the resulting star formation rate (SFR):
\begin{equation}
SFR=\sum_\mathrm{clumps} m_\mathrm{clump}/t_\mathrm{ff}
\end{equation}
where $t_\mathrm{ff}$ is the local free-fall time of the clump, calculated with the mass-weighted average density.  Only clumps with masses $>0.1\,$M$_\odot$ are included in the SFR calculation.  We calculate the SFR with smaller, separate clumps rather than with larger, aggregated clumps to better probe gravitational instability.  Because these are the most central, most dense, and most unstable parts of clumps resulting from the wind and collapse, we calculate the SFR with complete efficiency.  We integrate this SFR over the duration the clumps are gravitationally unstable for the final stellar mass created, which we tabulate in Table 2.  We note that we do not remove mass from the simulation through simulation.  As a result, the star formation rates we calculate may just be indicative of a trend of star formation.

\section{Results}

In all the simulations, we find the wind drives shock waves into the sphere, followed by a very strong, general compression due to ram pressure and thermal pressure of the wind and post-shock region.  However, depending on the parameters of the simulation, specifically the ram pressure, we find that the cloud can be ablated before it has the chance to collapse.  The simulations without collapse have KH times of about 20 kyr or shorter, meaning that the surface instabilities grow quickly and the cloud is stripped before gravity takes over.  In the other simulations, however, the opposite is true: cloud density increases dramatically through the initial compression, followed by gravitational collapse before KH instabilities or the perturbations in the wind can ablate the sphere.

\begin{figure*}
\includegraphics[width=0.9\paperwidth,angle=0]{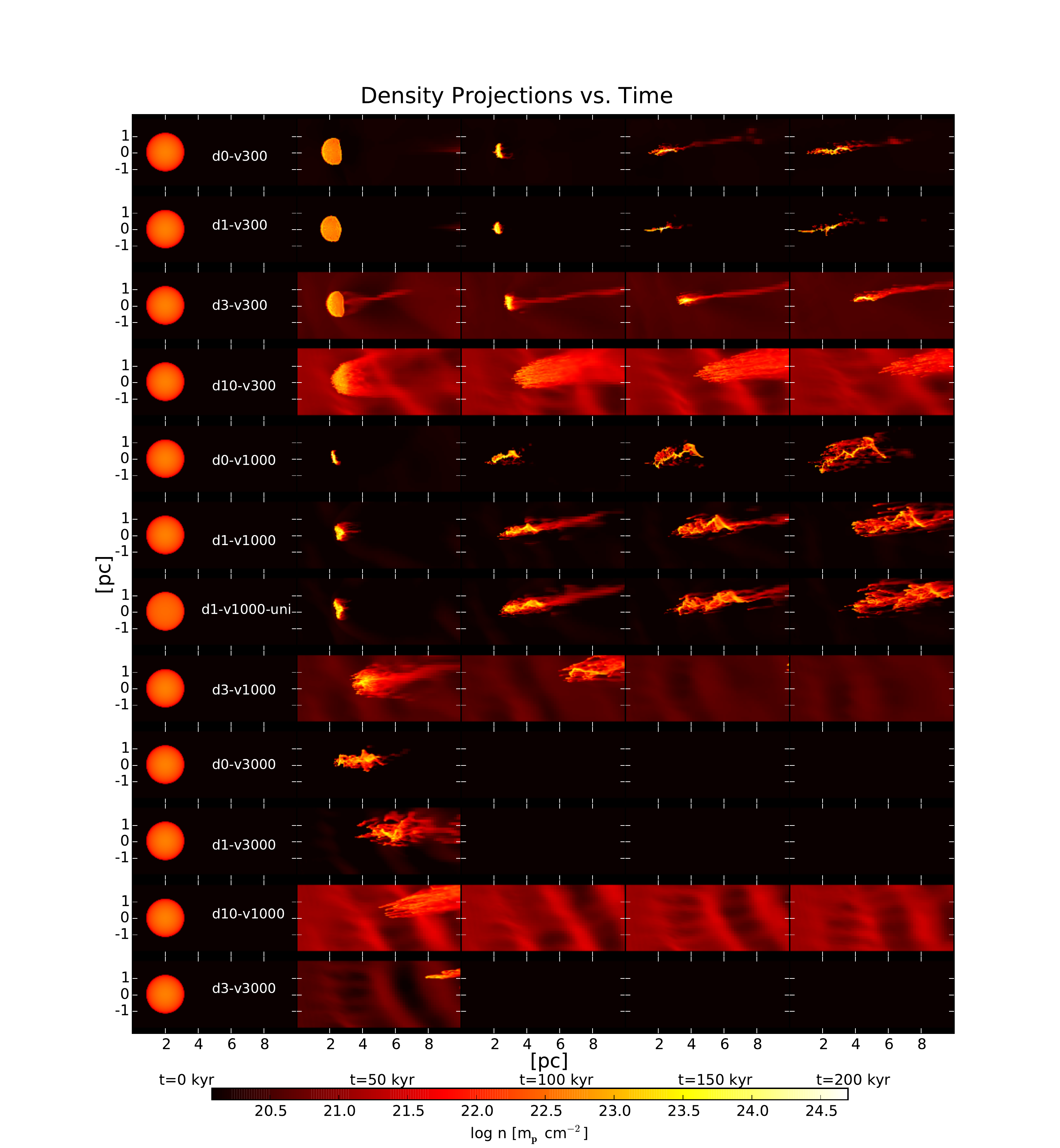} \\
\caption{Density projections for each simulation, shown in order of ram pressure from lowest to highest.}
\label{fig:density_projections}
\end{figure*}

In all of the simulations with ram pressures smaller than in d3-v1000, the wind compresses the sphere to form dense filaments in the direction of the incoming wind with the exception of d10-v300.  Figure \ref{fig:density_projections} shows the column densities of all twelve simulations in ascending order of ram pressure.  In the simulations with dense filaments, we see these filaments accelerated across the box faster with the increasing ram pressure.  The highest density parts of these filaments are toward the back end of the filament away from the incoming wind and shielded by the slightly less dense front part of the filament.  This is important for subsequent star formation because it prevents the shock from driving internal (small-scale) perturbations into the filament and clumps that would prevent gravitational collapse.  The exception simulation, d10-v300, has an incoming wind density of 10 \mpccm, and the spatially correlated perturbations cause this initial density to grow to over 100 $\mpccm$, and these dense perturbations ablate the sphere rapidly.  

\begin{figure*}
\begin{center}
\includegraphics[width=0.9\paperwidth,angle=0]{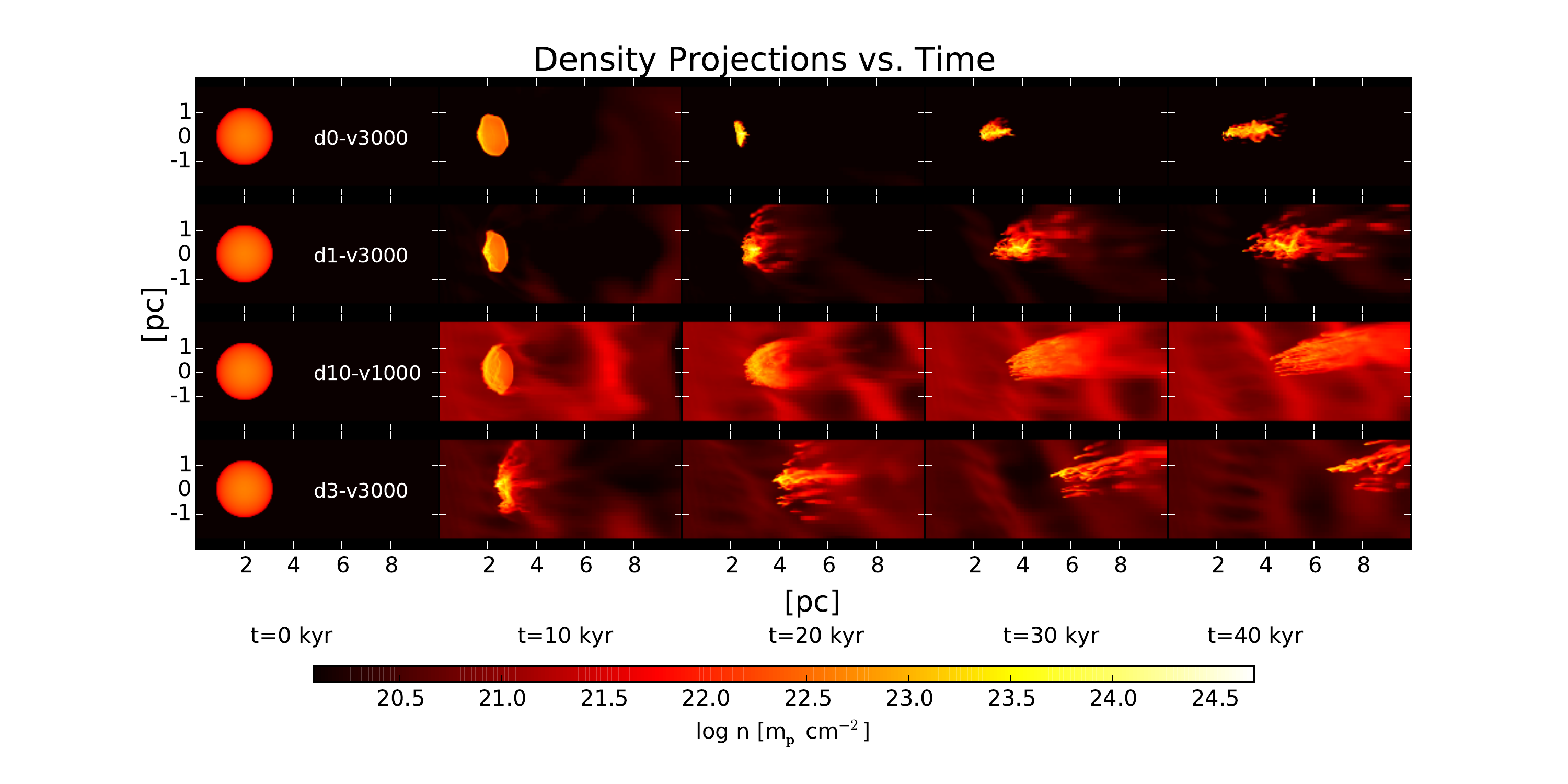} \\
\caption{Density projections for the four simulations with the highest ram pressure with shorter time intervals, shown in order of ram pressure from lowest to highest.}
\end{center}
\label{fig:density_projections_fast}
\end{figure*}

Figure \ref{fig:density_projections_fast} shows density projections for the four simulations with the highest ram pressures with smaller time intervals to resolve the rapid evolution of the sphere under those circumstances.  Run d0-v3000 evolves in a manner similar to d0-v1000 and d1-v1000, forming a dense, peanut shaped filament.  On the other hand, Runs d1-v3000, d3-v1000, and d3-v3000 all show rapid cloud ablation without the formation of dense filaments stretched out in the direction of the wind velocity.  In our analysis of star formation, we see that these dense filamentary structures are the result of gravitational collapse out-pacing cloud ablation and are critical to forming stars. 

\begin{figure*}
\includegraphics[width=0.9\paperwidth,angle=0]{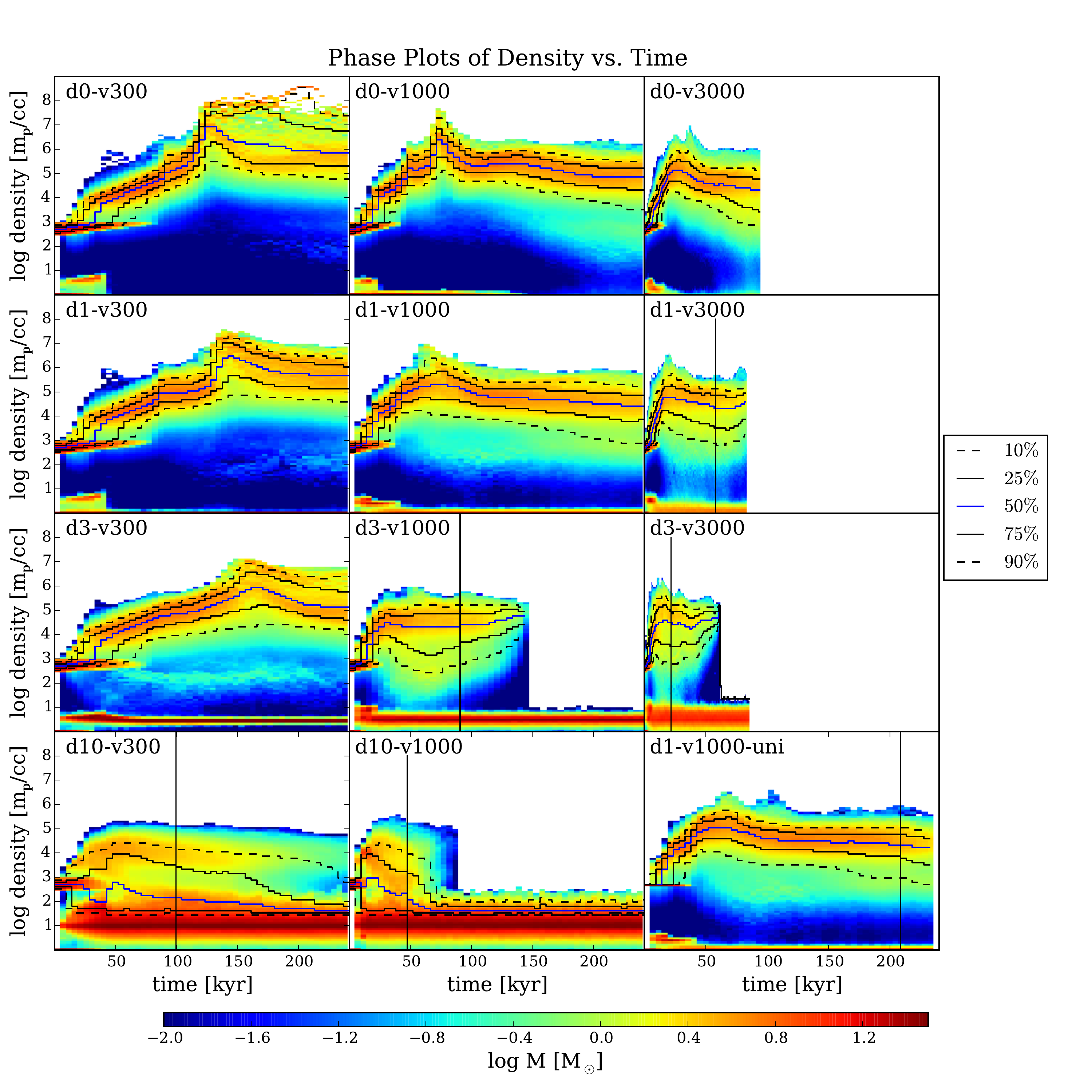} \\
\caption{Mass-weighted phase plots of density vs. time for all 12 simulations.  Percentile lines of mass are also plotted.  The black vertical line in six of the simulations marks the time when approximately 20\% of the dense gas ($\rho>150$ $\mpccm$) has been forced out the far end of the box.}
\label{fig:rho_time_phase}
\end{figure*}

Generally, as the ram pressure and momenta of the wind increase, the cloud crushing time and the time-to-maximum density decrease.  Figure \ref{fig:rho_time_phase} is a mass-weighted phase plot of density  density versus time for all 12 simulations.  In the 6 simulations with the highest ram pressure, a vertical black line is plotted, indicating the time when approximately 20\% of the dense gas has been forced out of the computational domain.  Runs d0-v1000, d1-v1000, d3-v1000, d0-v3000, d1-v3000, and d3-v3000 all show this effect.  Runs d0-v1000 and d1-v1000, both with velocities of 1,000 km s$^{-1}$, take 60 and 70 kyr to each peak densities of 4 and nearly 5 orders of magnitude greater than the initial peak density.  All the simulations with velocities of 3,000 km s$^{-1}$, Runs d0-v3000, d1-v3000, and d3-v3000, take 15, 25, and 35 kyr to maximum densities roughly 3.5 orders of magnitude greater than the initial maximum density.  However, Runs d0-v300, d1-v300, and d3-v300, all of which have a wind velocity of 300 km s$^{-1}$, take roughly 150, 160, and 170 kyr respectively to reach peak density, climbing 4-5 orders of magnitude.  In these three simulations with wind velocities of 300 km s$^{-1}$, as the ram pressure and momenta of the incoming winds increase, the time to peak density is slightly decreased, likely because of a small delay to the point where gravitational collapse can take over.  Runs d0-v300, d1-v300, and d3-v300 show by far the most gravitational collapse of any of the simulations.  

The two simulations with wind densities of 10 $\mpccm$, Runs d10-v300 and d10-v1000, reach peak densities of only two orders of magnitude greater than the initial value, and they do so quickly relative to the other simulations with the same velocity.  This is because the spatially correlated perturbations with densities starting at 10 $\mpccm$ can quickly grow to 100-150 $\mpccm$ with gravity, and these very dense perturbations have an extremely destructive impact, quickly ablating the cloud before any collapse can occur.  

Despite the similar density evolutions and morphological evolutions of the uniform sphere in Run d1-v1000-uni and the BE sphere of the same mass in Run d1-v1000, both with wind velocities of 1,000 km s$^{-1}$ and densities of 1 $\mpccm$, the two runs have considerably different star formation, demonstrating the importance of the gas distribution within the cloud.  The uniform sphere is already in an unstable state, we believe the different density structure, specifically higher densities at outer edges, better shield the inner core, allowing for earlier star formation.  

\begin{figure*}
\includegraphics[width=0.9\paperwidth,angle=0]{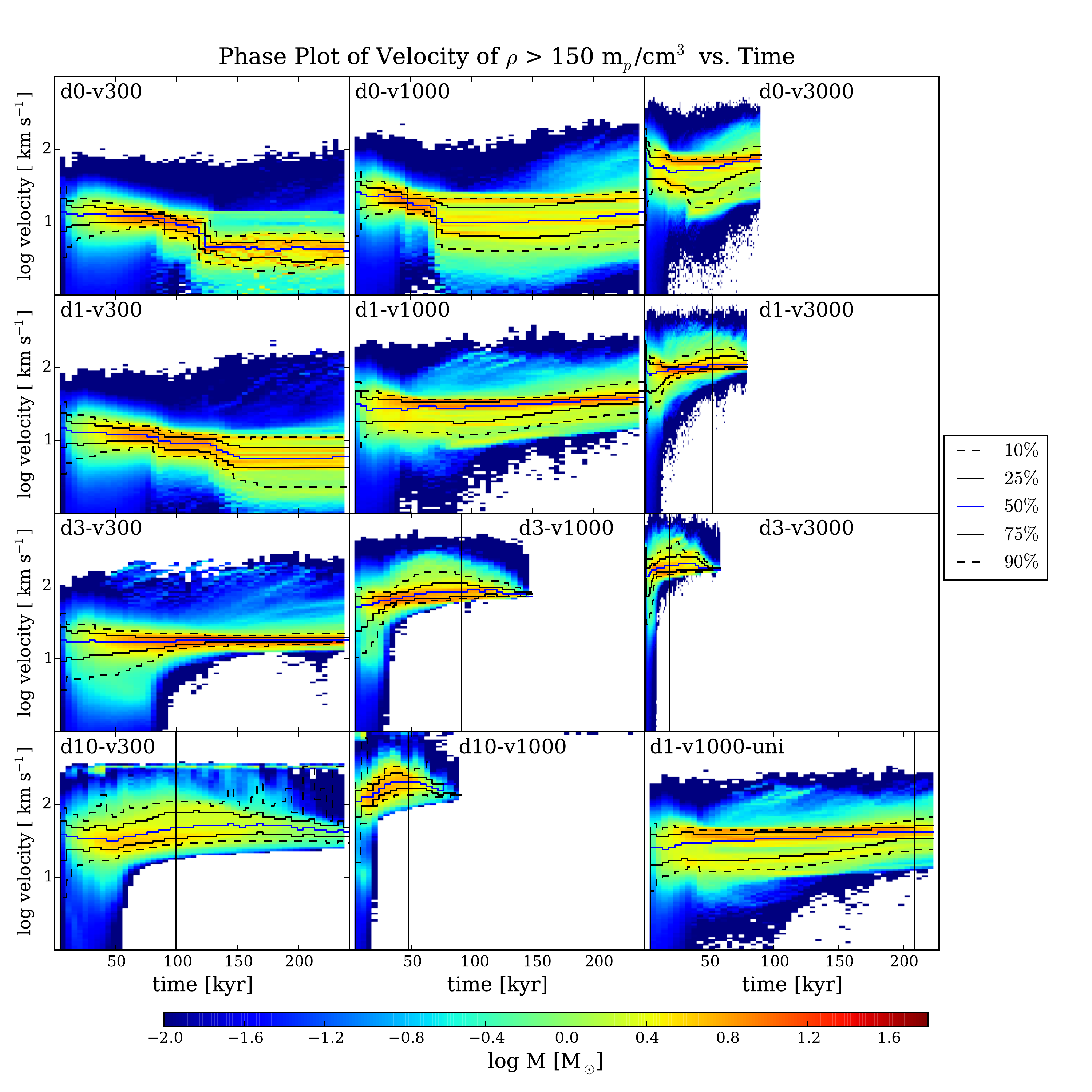} \\
\caption{Mass-weighted phase plots of velocity of dense gas ($\rho>150$$\mpccm$) vs. time for all 12 simulations.  Percentile lines of mass are also plotted.  The black vertical line in six of the simulations marks the time when approximately 20\% of the dense gas ($\rho>150$ $\mpccm$) has been forced out the far end of the box.}
\label{fig:vel_time_phase}
\end{figure*}

The dense gas inside the sphere is accelerated to a variety of velocities depending on the simulation parameters, but in all the runs, the entire sphere is initially accelerated to a velocity of at least 10 km s$^{-1}$.  Figure \ref{fig:vel_time_phase} shows the evolution of the velocity of all gas with densities greater than 150 $\mpccm$, basically limiting these plots to the gas in the sphere, dense filaments, or gas that has just been stripped.  On each row, as one looks left to right, the velocity of the wind is increased by a factor of roughly 3, and the velocity of the sphere is likewise increased by a factor of roughly 3.  In Run d0-v3000, more than 50\% of the dense gas has a velocity of $\sim$100 km s$^{-1}$ very early on.  All the runs with a ram pressure and momentum greater than or equal to that in d0-v1000 show continuous or late acceleration of the dense gas.    
  
 The differences between Run d1-v1000 and d3-v1000 and between Run d1-v3000 and d3-v3000 indicate how impactful the increase to a wind density of 3 $\mpccm$ is; all of the percentile lines show that the gas reaches much higher velocities, as much as 50 km s$^{-1}$ or more, far more quickly for the high wind density runs.  
 
 In Runs d10-v1000 and d10-v300, the wind density is 10 $\mpccm$, and this density combined with the spatially correlated velocity perturbations in gravity results in lumps of gas with densities of around 100 $\mpccm$ traveling at high velocities.  The high density, high velocity gas in the wind is represented by the top line feature of velocity 300 km $s^{-1}$ in the phase plot for Run d10-v300, and the top left feature of velocity $\sim$1,000 km $s^{-1}$ Run d10-v1000.  All of this gas has a density of at least 150 $\mpccm$.  This high velocity high density gas in the wind quickly ablates the sphere, preventing it from reaching high densities or velocities, as well as preventing any collapse.  
 
In Runs d0-v300, d1-v300, and d0-v1000, all with low ram pressure and momentum, the sphere is initially accelerated, but as it collapses its density increases and volume decreases, resulting in a subsequent velocity decrease.  
 
Here, Equation \ref{eq:wind} provides an interesting point of comparison.  Without ablation or collapse, uniform spheres would theoretically be accelerated to velocities similar to what we see in our simulations, with some differences.  For example, in Run d0-v300 with the lowest ram pressure, the analytical model and our simulations both show velocities of 1 to tens of km s$^{-1}$.   However, in the analytical model, the velocity continually increases.  In the simulation, the median velocity is initially high as the sphere is compressed, then decreases as gravitational collapse takes over and the densities increase by several orders of magnitude.  The same happens in d0-v1000 and d1-v300.  On the higher end of the velocity spectrum with wind velocities of 3,000 km s$^{-1}$, the cloud never achieves the velocities in the simulations that the analytical models predict, not even the fastest 10\% of the cloud.  The median velocities of the clouds in runs d0-v3000, d1-v3000, and d3-v3000, are around 100 km s$^{-1}$, whereas the analytic models predict velocities closer to 1,000 km s$^{-1}$.  The reason for this difference is that besides accelerating it, the wind also strongly changes the cloud properties particularly by decreasing its effective cross section by compressing it.  Additionally, the cloud ablation time is much shorter than the time it would take to reach these velocities ($\sim$100 kyr).  When the values for the cloud radius in Equation 17 are decreased by factors of 2-10 and the density increases by 2-4 orders of magnitude, as the cloud collapses into dense filaments, the expected values for the cloud velocity are likewise decreased to values under 100 km s$^{-1}$.  Overall, the simulations show that the fluid dynamics of non-uniform spheres experiencing turbulent winds behave differently than simple estimates may predict because of the evolving properties of the cloud with time.  

\begin{figure*}
\includegraphics[width=0.9\paperwidth,angle=0]{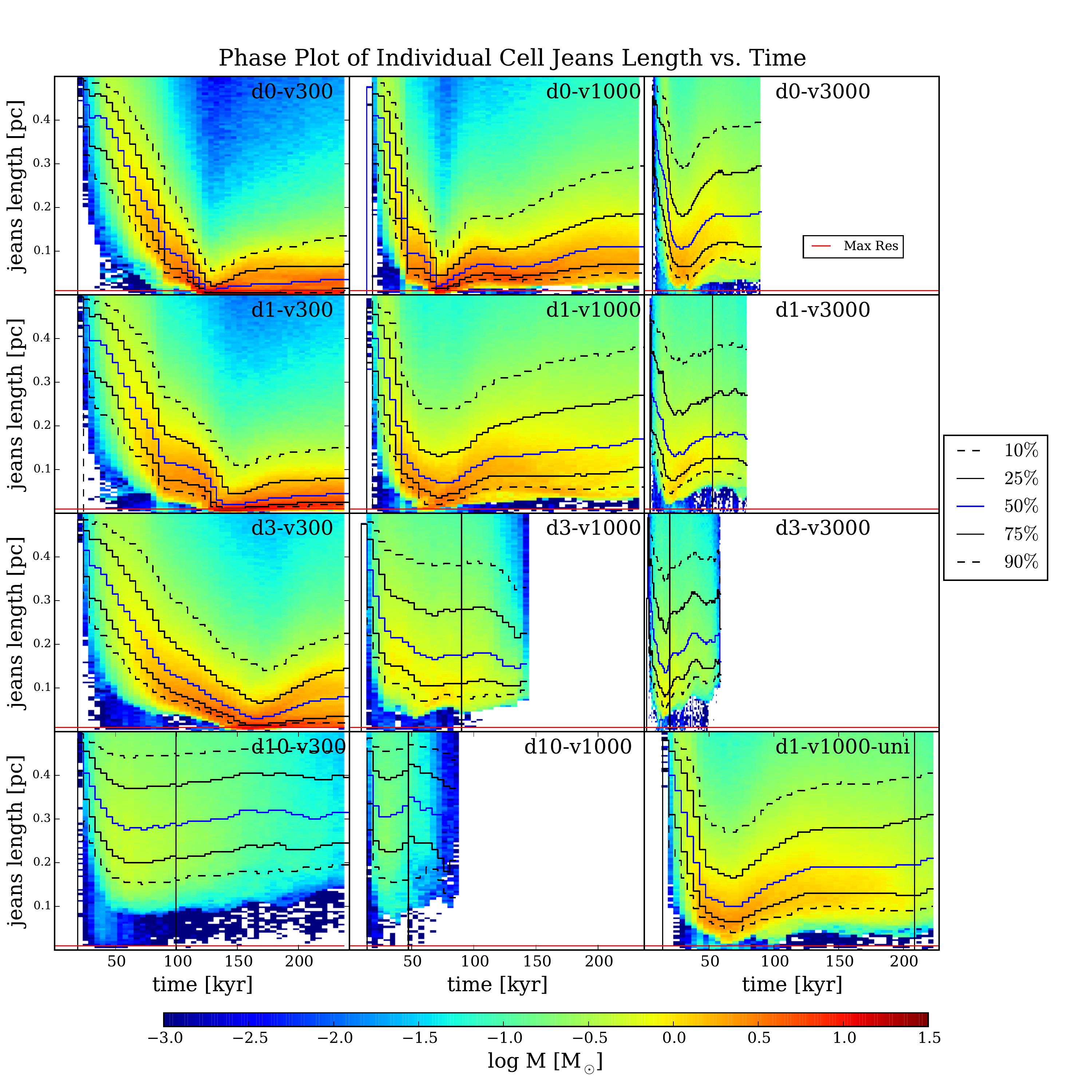} \\
\caption{Mass-weighted phase plots of jeans length of individual cells vs. time for all 12 simulations.  Run label, wind density, and wind velocity are printed in each plot for clarity.  Percentile lines of mass are also plotted.  The black vertical line in six of the simulations marks the time when approximately 20\% of the dense gas ($\rho>150$ $\mpccm$) has been forced out the far end of the box.}
\label{fig:jeans_time_phase}
\end{figure*}

As the dense filaments form and the gas is accelerated across the box, gravitational collapse can occur on the scales as large as a clump and as small as our maximum resolution.  Figure \ref{fig:jeans_time_phase} shows the mass-weighted phase plot of the Jeans length of the individual cells vs. time for each of the 12 simulations.  We over-plot a red line at 0.01 pc, representing the maximum resolution of our simulation.  With Adaptive Mesh Refinement (AMR) codes, this also represents the maximum resolution of gravity, meaning that once the smallest cells obtain a particular density, gravity within each cell can no longer be resolved.  This means that gravitational collapse could occur inside the smallest cell without the code being able to resolve it.  In turn, any gravitational collapse we calculate on a larger scale, is therefore a lower limit on gravitational collapse.  With better spatial resolution, we could likely detect more collapse, deeper potentials, and more subsequent star formation.  
 
 Figure \ref{fig:jeans_time_phase} shows how much mass exists in the maximally resolved cells that are too dense to resolve the Jeans length.  For the most part, this occurs because of the compression phase of each simulation.  Runs d0-v1000, d1-v1000, d0-v3000, and d1-v3000 show this effect the most.  In these simulations, after the clumps reach their peak densities, push back from growing internal pressures overcomes the gravitational potential and moves the gas out of the regime where it would collapse inside a single cell.  This figure shows, however, that for the most part, the potential within each cell is reasonably well resolved.

\subsection{Clump Analysis}
\begin{figure*}
\includegraphics[width=0.9\paperwidth,angle=0]{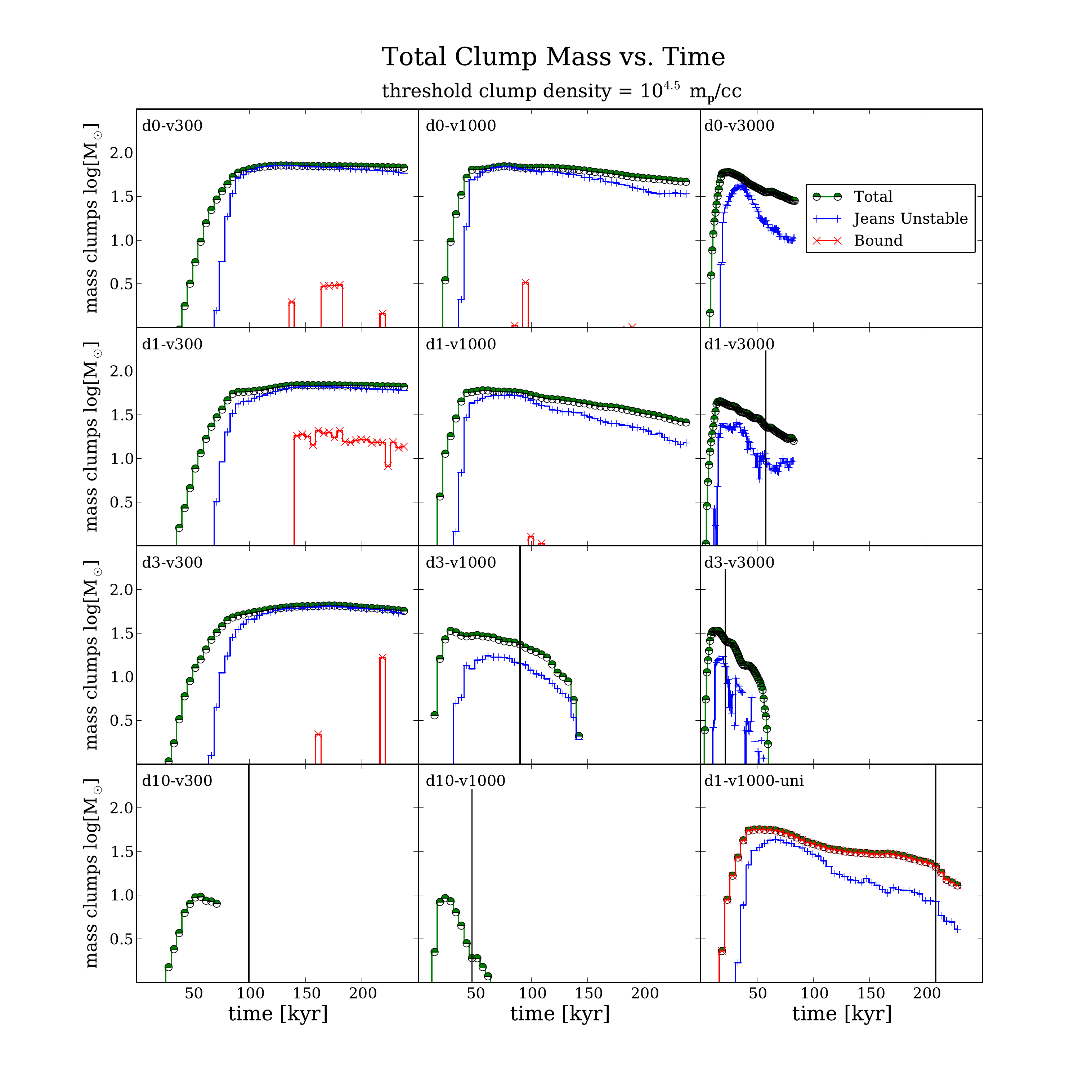} \\
\caption{Total clump mass with a threshold density of 10$^{4.5}$ $\mpccm$, Jeans unstable clump mass, and bound clump mass vs. time for all 12 simulations.  Run label, wind density, and wind velocity are printed in each plot for clarity.  Percentile lines of mass are also plotted.  The black vertical line in six of the simulations marks the time when approximately 20\% of the dense gas ($\rho>150$ $\mpccm$) has been forced out the far end of the box.}
\label{fig:clump_mass_time_low_thresh}
\end{figure*}

\begin{figure*}
\includegraphics[width=0.9\paperwidth,angle=0]{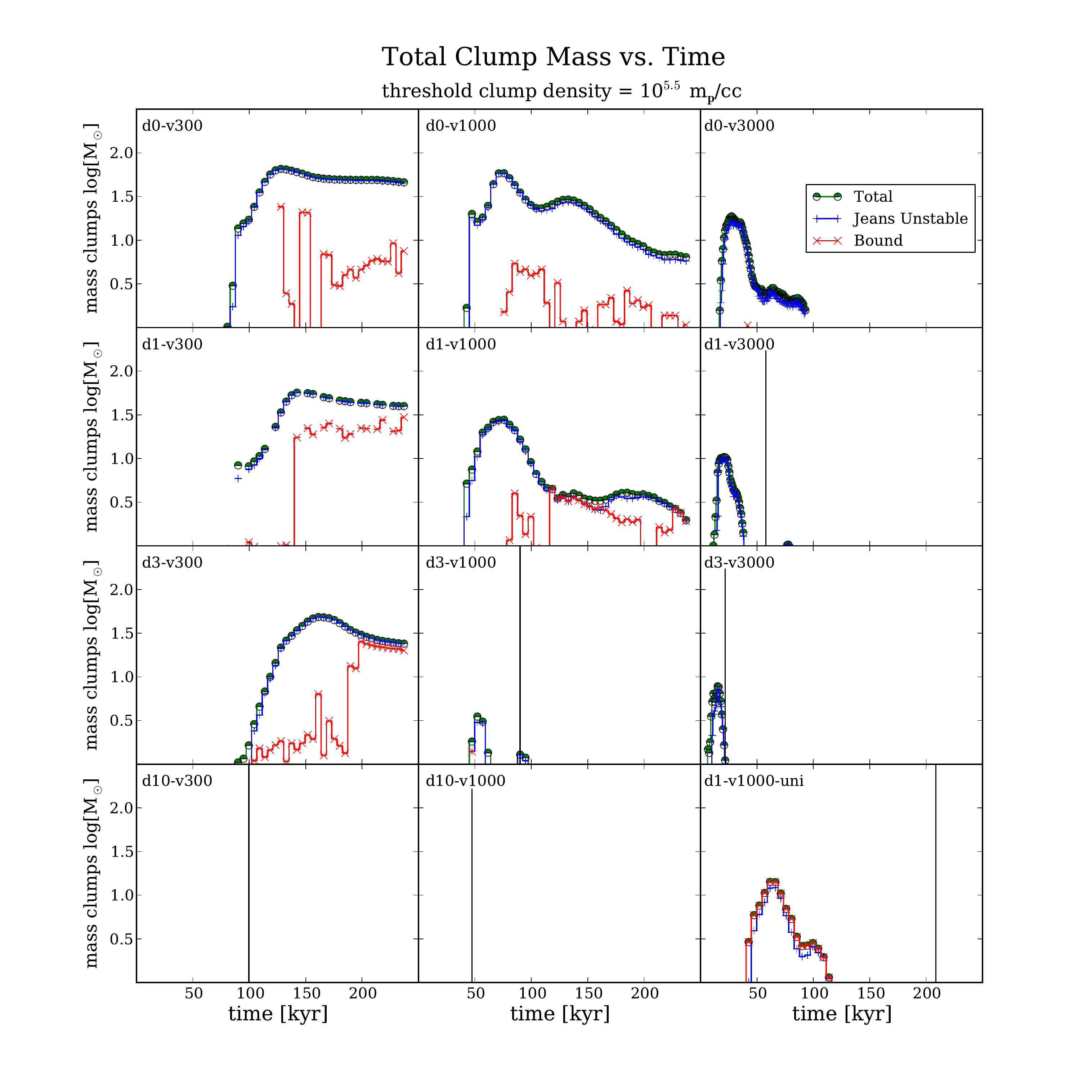} \\
\caption{Total clump mass with a threshold density of 10$^{5.5}$ $\mpccm$, Jeans unstable clump mass, and bound clump mass vs. time for all 12 simulations.  Run label, wind density, and wind velocity are printed in each plot for clarity.  Percentile lines of mass are also plotted.  The black vertical line in six of the simulations marks the time when approximately 20\% of the dense gas ($\rho>150$ $\mpccm$) has been forced out the far end of the box.}
\label{fig:clump_mass_time_high_thresh}
\end{figure*}

To investigate gravitational collapse and subsequent star formation, we evaluate the gravitational stability or instability of the individual clumps that form in the dense filaments.  Again, to best probe gravitational instability, we employ a threshold density of 10$^{5.5}$ $\mpccm$ and a peak-to-saddle ratio of 1.2 to identify the clumps most likely to collapse.  However, to show that these are indeed the dense central cores, we also look at clumps with threshold densities of 10$^{4.5}$ $\mpccm$ so that we can properly analyze the more diffuse envelopes.  Figure \ref{fig:clump_mass_time_low_thresh} shows the evolution of total clump mass versus time with the lower threshold of 10$^{4.5}$ $\mpccm$ for all twelve simulations.  Figure \ref{fig:clump_mass_time_high_thresh} shows the evolution of total clump mass versus time with the higher threshold of 10$^{5.5}$ $\mpccm$.  The discrepancy in mass but similarity in evolution of the two plots together show that the clumps with the higher density threshold are truly the dense cores of more massive, larger clumps and thus are shielded from the turbulent gas outside the larger clump.

Figure \ref{fig:clump_mass_time_high_thresh} shows that only those simulations with velocities of 300 and 1,000 km s$^{-1}$ and densities of 0.5, 1, and 3 \mpccm form the dense clumps where stars form.  The green line shows the sum of mass of all of the clumps with masses $>0.1$M$_\odot$, the blue shows the total mass of the Jeans-unstable clumps, and the red shows the total mass of gravitationally bound and Jeans-unstable gas that will form stars.  

None of the simulations with the wind velocity of 3,000 km s$^{-1}$ form stars.  None of the simulations with a wind density of 10 $\mpccm$ form stars because the velocity perturbations can cause the wind density to climb to up to 150 $\mpccm$, rapidly ablating the sphere.  The simulation in which we have a uniform sphere rather than a BE sphere forms stars earlier as of result of the dense edges of the sphere, indicating that the distribution of gas does in fact make a difference in potential star formation.

The black vertical line in six of the simulations marks the time when approximately 20\% of the dense gas ($\rho>150$ $\mpccm$) has been forced out the far end of the box.  These lines show that in those simulations with the highest ram pressure, the clump analysis is valid because in those simulations, the clumps are stripped and destroyed before 20\% of the dense gas has been forced out.  Therefore we expect that no star formation would occur in these simulations, even if we were able to track the simulations longer.  

Furthermore, we see that nearly all of the gas in the dense clumps is Jeans-unstable, indicating that the internal (small-scale) velocities of the clump rather than the thermal pressure supports the clump.  This is significant because the wind and subsequent turbulence from the AGN feedback transfers a significant amount of kinetic energy to the clump, including to the internal velocities.  However, as we focus on the higher density threshold clumps which comprise the cores of the lower density threshold clumps, we see that part of the clump that is shielded from the surrounding turbulence, and lacks the high internal velocities to support the clump against gravitational collapse.

\begin{figure}
\includegraphics[width=1.\linewidth,angle=0]{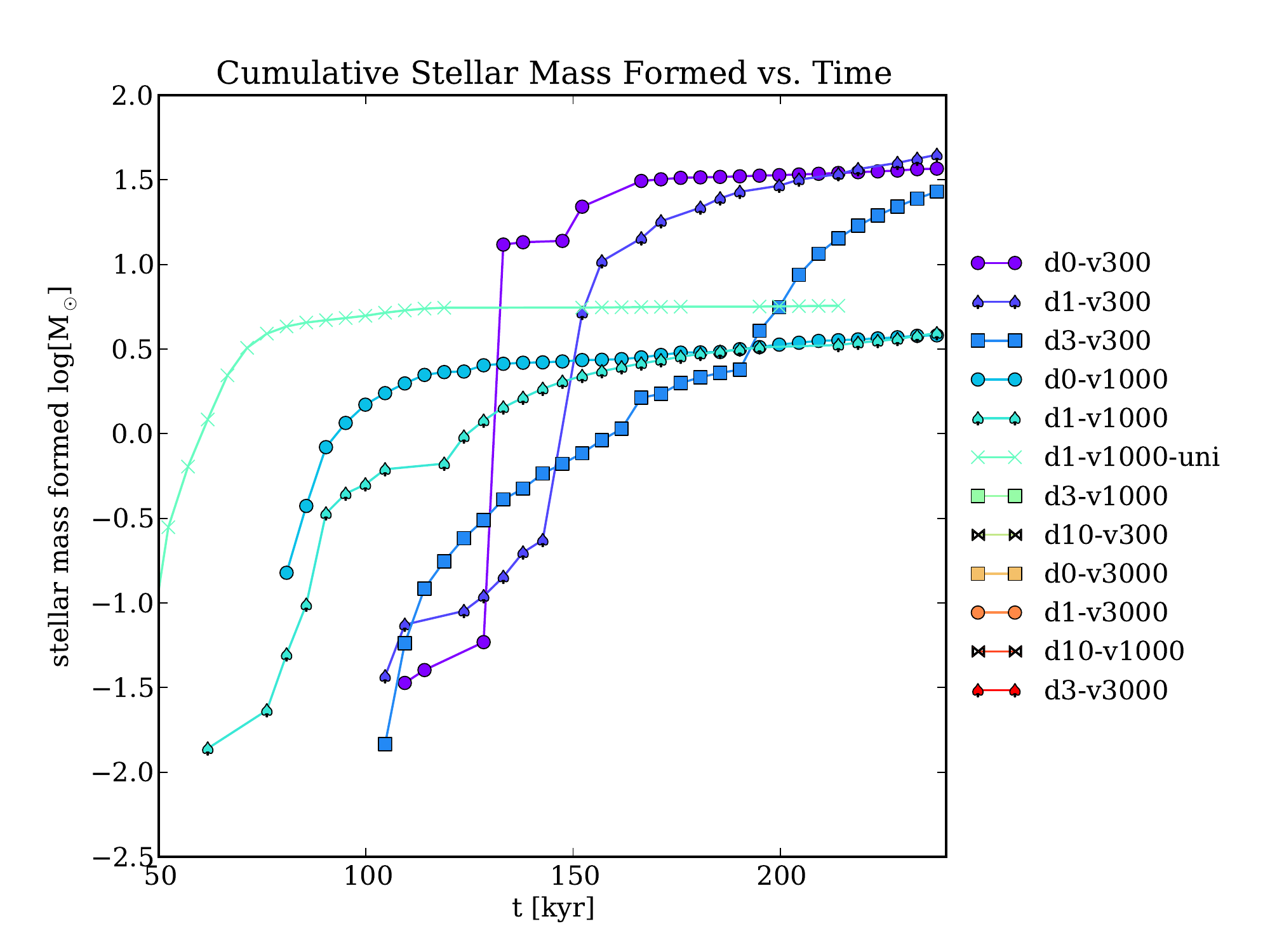} \\
\caption{Cumulative stellar mass formed vs. time.  Total mass of stars formed presented in Table 2.}
\label{fig:star_massed_formed}
\end{figure}

As clumps collapse, we calculate the resulting cumulative stellar mass formed using the method described in Section 3.5.  We find that with increasing ram pressure, stars are formed earlier, and less stellar mass is formed.  Figure \ref{fig:star_massed_formed} shows the cumulative stellar mass formed vs. time in all the simulations, shown in the order of ram pressure.  The simulations with wind velocities of 1,000 km s$^{-1}$ begin forming stars earlier, around 50 kyr into the simulations as the BE sphere is compressed to much higher densities.  However, the SFR in these runs decreases with time because the clump mass decreases with time as they are stripped.  The SFR increases with time in the simulations with wind velocities of 300 km s$^{-1}$ because the peak densities in these simulations tend to increase steadily through the simulation, and again, the resulting star formation rate mirrors the same pattern.  

An important result of the study is that we find with increased ram pressure in the wind there is a decrease in the mass of stars formed, leading up to a threshold value of wind ram pressure of 1.67e-8 dyn cm$^{-2}$, wind momentum of 3.34e-16 g cm s$^{-1}$.  Table \ref{tab:Table_2} shows the final mass of stars formed in each of the simulations.  These simulations are ordered by the ram pressures of their respective winds, roughly the same order as their respective momenta.  

Star clumps from the simulations with higher ram pressure clearly have higher velocities of up to 30 km s$^{-1}$, also shown in Table \ref{tab:Table_2}.  These velocities are consistent with what Equation 14 would predict (tens of km s$^{-1}$).  These results are also consistent with the positive radial velocity boost of stars formed in galaxies with AGN from theoretical work by \citet{Dugan14}, \citet{Gaibler12}, and \citet{Silk12}; however, we find smaller velocities here than in those studies.  We believe that a sphere larger than the one we simulate would survive longer and therefore have more time to be accelerated to higher velocities by the shock, reaching velocities closer to those calculated in the aforementioned studies.

\begin{deluxetable*}{cccccccccc} 
\tablecolumns{10}
\tablecaption{Stars Formed}
\tablehead{ \colhead{Run} & \colhead{$\rho$} & \colhead{$v_w^{(a)}$} & \colhead{P$_\mathrm{ram}^{(b)}$ } & \colhead{p$^{(c)}$} & \colhead{$\mathcal{M}^{(d)}$} & \colhead{$t_\mathrm{cc}^{(e)}$} & \colhead{$t_\mathrm{KH}^{(f)}$ } & \colhead{SF$^{(g)}$} & \colhead{$\bar{v}^{(h)}$} \\
\colhead{Label} & \colhead{[$\mpccm$]} & \colhead{[km s$^{-1}$]} & \colhead{[dyne cm$^{-2}$]} & \colhead{ [g cm} & \colhead{} & \colhead{[kyr]} & \colhead{[kyr]} & \colhead{M$_\odot$} & \colhead{[km} \\
\colhead{} & \colhead{} & \colhead{} & \colhead{} & \colhead{s$^{-1}$]} & \colhead{} & \colhead{} & \colhead{} & \colhead{} & \colhead{s$^{-1}$]}  }
d0-v300 & 0.5 & 300 & 7.52e-10 & 5.01e-17 & 1.07e+03 & 95.84 & 166.09 &  37.35  & 3.20\\  
d1-v300 & 1.0 & 300 & 1.50e-09 & 1.00e-16 & 1.07e+03 & 67.77 & 117.51 &  47.89 & 5.09 \\  
d3-v300 & 3.0 & 300 & 4.51e-09 & 3.01e-16 & 1.07e+03 & 39.13 & 68.00 &  29.41 & 17.37 \\  
d0-v1000 & 0.5 & 1,000 & 8.35e-09 & 1.67e-16 & 3.57e+03 & 28.75 & 49.83 &  3.85  & 9.40\\ 
d10-v300 & 10.0 & 300 & 1.50e-08 & 1.00e-15 & 1.07e+03 & 21.43 & 37.53 &  0.00 & -- \\ 
d1-v1000 & 1.0 & 1,000 & 1.67e-08 & 3.34e-16 & 3.57e+03 & 20.33 & 35.25 &  4.03  & 33.28 \\  
\hline
d1-v1000-uni & 1.0 & 1,000 & 1.67e-08 & 3.34e-16 & 3.57e+03 & 20.33 & 24.54 & 5.71 & 14.08 \\
\hline
d3-v1000 & 3.0 & 1,000 & 5.01e-08 & 1.00e-15 & 3.57e+03 & 11.74 & 20.40 &  0.09   & --\\  
d0-v3000 & 0.5 & 3,000 & 7.52e-08 & 5.01e-16 & 1.07e+04 & 9.58 & 16.61 &  0.00   & --\\  
d1-v3000 & 1.0 & 3,000 & 1.50e-07 & 1.00e-15 & 1.07e+04 & 6.78 & 11.75 &  0.00   & --\\  
d10-v1000 & 10.0 & 1,000 & 1.67e-07 & 3.34e-15 & 3.57e+03 & 6.43 & 11.26 &  0.00   & --\\  
d3-v3000 & 3.0 & 3,000 & 4.51e-07 & 3.01e-15 & 1.07e+04 & 3.91 & 6.80 &  0.00   & --  
\enddata
\tablenotetext{}{\begin{center}Run d1-v1000-uni: uniform sphere with the same mass and radius as the BE sphere.\\
$^{(a)}$ Wind velocity.\\
$^{(b)}$ Ram pressure of wind.\\
$^{(c)}$ Wind momentum.\\
$^{(d)}$ Mach number- wind velocity relative to sound speed of sphere.  \\
$^{(e)}$ Cloud crushing time.\\
$^{(f)}$ Kelvin-Helmholtz time.\\ 
$^{(g)}$ Cumulative stellar mass formed.  \\
$^{(h)}$ Mass-weighted average velocity of star forming clumps at the end of the simulation.  \\
\end{center}
}
\label{tab:Table_2}
\end{deluxetable*}

\section{Discussion}
These simulations cover the broad range of physical parameters for the winds caused by AGN feedback from jets or radio quiet quasars.  
These winds can drive the peak densities of the cloud by up to five orders in magnitude through the formation of dense filaments and clumps, subsequently resulting in a drastic shortening of the free fall time.  We find threshold values for the ram pressure and momentum of the incoming wind below which the resulting clumps' binding energy overcomes internal kinetic energy and pressure, resulting in gravitational collapse and star formation (shown in Table \ref{tab:Table_2}).  Increased ram pressure and momentum of the incoming wind decrease the total mass of resulting stars but increase the velocities of formed stars, some achieving velocities of up to 35 km s$^{-1}$.  

The results require some interpretation to be applicable to the debate between positive and negative feedback.  We have taken a dense cloud of gas, that by itself would not collapse, and then applied realistic physical circumstances expected for AGN feedback that induce star formation in half of the simulations.  In this sense, our results are complementary to previous studies that shock uniform spheres, that would collapse on their own, preventing the spheres from collapsing and showing negative feedback.  However, nearly all of those do not include gravity and shock uniform spheres.  In our suite of simulations, we include gravity and simulate BE spheres, and find that the distribution of gas within the sphere can make a considerable difference to the results.

Our simulations show the formation of dense filaments reported in \citet{Pittard10}, \citet{Orlando05}, \citet{Mellema02}, \citet{Cooper09}, and \citet{Zubovas14_pressure}.  Our results are in agreement with the latter three studies in that the geometry of the cloud impacts the results.  Only \citet{Zubovas14_pressure} includes gravity, and we both report star formation.  However, we build on their results by identifying the velocity, density, and ram pressure range in which star formation is possible.  We also include substantially higher velocities, characteristic of AGN jet cocoons and outflow bubbles.  With respect to the three phases of uniform cloud shocking introduced by \citet{Klein94} and identified in \citet{Mellema02}, among others, \citet{Cooper09} includes a fourth phase in which the cloud is ablated and destroyed.  We find another possible phase in which gravitational collapse outpaces ablation.  

Comparisons between cloud collapse time and ablation time have been made previously.  \citet{Wagner11} argued that for clouds less than 50 pc in size the collapse time would be longer, resulting in negative feedback.  However, \citet{Gaibler12} and \citet{Mellema02} speculated that only the outer, more diffuse envelopes of these clouds might be ablated, leaving behind the dense inner core to continue collapsing.  Our results corroborate this idea.  We find that a stratified density distribution of the sphere may change the manner in which the cloud is stripped, the mass that is lost, and how the remaining core is shielded and collapses, as in \citet{Nakamura06}.  With a uniform sphere, shielding from the sphere's dense edge prevents the wind from injecting kinetic energy into the sphere, which leaves the thermal pressure as the main mechanism of support.  This is considerably different from the other simulations, in which the internal velocity dispersions are crucial to the clump's support.  It is also worth noting that the extreme ram pressures resulting from AGN feedback quickly change the nature of the cloud being shocked, increasing the density and ablation time, decreasing the free fall time, and causing star formation.  In the longer term, however, after the outer envelopes have been stripped and the cores have collapsed to form stars, the remaining gas budget for continued star formation may be depleted, causing long term negative feedback after a burst of positive feedback. Strong fragmentation of the cloud may result in cloudlets too small too collapse or too small to form stellar objects.

Within an entire galaxy, as a jet cocoon or outflow bubble expands through the galaxy, some clouds will be struck with velocities of 3,000 km s$^{-1}$ while others, shielded by other clouds, will be struck with slower velocities of 1,000 km s$^{-1}$ or even 300 km s$^{-1}$ depending on the location.  \citet{Wagner15} discusses the idea that positive and negative feedback can occur simultaneously within the same galaxy, depending on the size of the clouds being struck with shocks, that the larger clouds will collapse under the pressure and form stars while the smaller clumps will be blown away.  Since we did not change the cloud properties systematically in our study, we cannot assess the impact of variations in the typical sizes clouds or cloud complexes, e.g. with redshift \citep{Gaibler2014}.  We only vary the the winds striking the clouds, but we too find that simultaneous positive and negative feedback is possible in the same galaxy provided that the winds will depend on time and location within the galaxy.  We also include gravity and have spatial resolutions that are much better than those achievable in galaxy wide simulations. 

We see that the star formation efficiency decreases with ram pressure.  This is an interesting result that fits as a data point on the spectrum between negative and positive feedback, that AGN feedback can trigger star formation, but at a possibly lower efficiency than would otherwise occur, depending on the parameters of the feedback. There is, however, another intricacy that might be relevant: if star formation is not entirely prevented by high ram pressures, the time scale for collapse and the formation of stars within a cloud should be much shorter and happen more synchronously rather than be spread over the (long) free-fall time of the uncompressed cloud or cloud complex. Correspongingly, the impact of (negative) stellar feedback within the cloud by other young stars should be considerably smaller and thereby increase the efficiency of star formation.

Among other interesting conclusions, \citet{Cooper09} found that that the wind destroyed regions of lower density faster than those of higher density.  Combined with our results in which the density of the core clumps can climb up to five orders of magnitude in $\sim$50 kyr, this would indicate that the clumps would survive for long periods of time.  \citet{Cooper09} also found that radiative cooling increases the survival time of the clouds.  The clouds in \citet{Cooper09} survive up to 1 Myr, achieving velocities of 150--400 km s$^{-1}$, and it is reasonable to believe that our dense clumps could survive that long if they were not already collapsing.  Those velocities are higher than our clump velocities at the simulation's end, but that is to be expected because our simulations have maximum times of 250 kyr and because our clumps are much denser.

The results from our simulations are similar to our previous studies, \citet{Dugan14} and \citet{Gaibler12}, which show both that feedback in radio-loud galaxies may be positive and that stars formed during this period of feedback will have a positive velocity bias.  Specifically, these stars are typically formed with greater radial and vertical (off the disk) velocity distributions.  While we found velocities of up to 35 km s$^{-1}$ within the simulated time span for our small cloud, we note that larger clouds might survive longer, would be accelerated to higher velocities due to the larger effective cross section and thus form stars with higher peculiar velocities.  

In turn, this also provides further computational support to previously existing theory linking AGN feedback with hypervelocity stars \citep{Silk12}.  In this scenario, the cocoon or bow shock of a jet accelerates a pocket of gas to high velocities while compressing it to form stars.  \citet{Silk12} makes analytical arguments based on energy and momentum, but here we provide computational evidence to the same end.  Again, because our simulations are limited in the size of the cloud and the time with which we have to accelerate the cloud, we do not achieve stellar velocities as high as some of those observed in the Milky Way.  However, what we show here is that the phenomenon is possible on smaller scales.  

It would be interesting in the future to extend the simulations to larger gas complexes such as giant molecular clouds (GMC). For these, the assumption of having BE spheres is no longer suitable and a turbulent GMC would need to be included. This may address the idea that the larger the cloud is, the more positive the feedback could be.

\section{Conclusion}
Our simulations show that the unique combination and range of ram pressure and thermal pressure AGN feedback provides can either trigger or prevent star formation in dense clumps within the ISM, depending mostly on the ram pressure.  Our setup includes a broad range of parameters expected for winds originating from AGN jet cocoons or quasar winds. In the simulations that form stars, the wind compresses the cloud in the high-pressure post-shock region, then followed by the simultaneous stripping and collapse of the sphere.  The initial compression from the shock and not gravity causes the peak density in the sphere to climb by as much as 5 orders of magnitude in as little as 50 kyr, reaching regimes where gravitational collapse overcomes internal pressure and kinetic energy.  During this phase, the free fall time of the cloud drops from 3 Myr initially, to as low as 10 kyr in subsequent clumps.

We analyze these clumps resulting from this collapse, and subsequent star formation within the clumps, finding indications of both positive and negative feedback, depending on the parameters of the wind.  We find that the stratified density profile of Bonnor--Ebert spheres can have a considerable impact on gravitational collapse and star formation.  We find that increased ram pressure, momentum, and Mach number of the incoming wind decreases the total mass of stars formed, but increases the velocities of these stars. Furthermore, star formation would happen on a much shorter time scale and more synchronously.  We identify threshold values for ram pressure and momentum of $\sim 2\times10^{-8}$ dyn cm$^{-2}$ above which star formation is not expected to occur because high internal velocities generated within the clumps support them against gravitational collapse long enough for the cloud to be destroyed.    

Our results indicate that simultaneous positive and negative feedback will be possible in a single galaxy as AGN wind parameters will vary with location within a galaxy.  Similar phenomena are observed in AGN \citep{Cresci15,Cresci15a,Carniani16}.

\acknowledgments

ZD was supported by a Centre for Cosmological Studies Balzan Fellowship.  
VG was supported by the Sonderforschungsbereich SFB 881 ``The Milky Way System'' (subproject B4) of the German Research Foundation (DFG).
The research of JS has been supported at IAP by  the ERC project  267117 (DARK) hosted by Universit\'e Pierre et Marie Curie - Paris 6   and at JHU by NSF grant OIA-1124403.  We would like to thank Alex Wagner for his input.  



\end{document}